\documentclass{aa}
\usepackage{amsmath}
\usepackage{amsfonts}
\usepackage{todo}
\usepackage{upgreek}
\usepackage[T1]{fontenc}
\usepackage{wasysym}
\usepackage[utf8]{inputenc}
\usepackage{lmodern}
\usepackage{ulem}
\usepackage{float}
\usepackage{tabularx} 

\graphicspath{ {files/} }

\begin{document}
%TC:ignore
%TC:endignore

\title{Asgard/NOTT: L-band nulling interferometry at the VLTI}
\subtitle{I. Simulating the expected high-contrast performance}
\author{Romain Laugier \inst{1}
        \and
        Denis Defrère \inst{1}
        \and
        Julien Woillez \inst{2}
        \and
        Benjamin Courtney-Barrer \inst{2,10}
        \and
        Felix A. Dannert \inst{3,4}
        \and
        Alexis Matter \inst{5}
        \and
        Colin Dandumont \inst{6} 
        \and
        Simon Gross \inst{7}
        \and
        Olivier Absil \inst{6}
        \and
        Azzurra Bigioli \inst{1}
        \and
        Germain Garreau \inst{1}
        \and
        Lucas Labadie \inst{9}
        \and
        Jérôme Loicq \inst{6,8}
        \and
        Marc-Antoine Martinod \inst{1}
        \and
        Alexandra Mazzoli \inst{6}
        \and
        Gert Raskin \inst{1}
        \and
        Ahmed Sanny \inst{7,9}
        }
\institute{
        Institute of Astronomy, KU Leuven, Celestijnenlaan 200D, 3001 Leuven, Belgium   %1
        \and
        European Southern Observatory Headquarters, Karl-Schwarzschild-Straße 2, 85748 Garching bei München, Germany  %2
        \and
        ETH Zurich, Institute for Particle Physics \& Astrophysics, Wolfgang-Pauli-Str. 27, 8093 Zurich, Switzerland  %3
        \and
        National Center of Competence in Research PlanetS (www.nccr-planets.ch)  %4
        \and
        Université Côte d'Azur, Observatoire de la Côte d'Azur, CNRS, Laboratoire Lagrange, France %5
        \and
        STAR Institute, University of Liège, 19C allée du Six Août, 4000 Liège, Belgium  %6
        \and
        MQ Photonics Research Centre, School of Mathematical and Physical Sciences, Macquarie University, NSW, 2109, Australia %7
        \and
        Faculty of aerospace Engineering, Delft University of Technology, 2629 HS Delft, the Netherlands  %8
        \and
        I. Physikalisches Institut, Universität zu Köln, Zülpicher Str. 77, 50937 Köln, Germany  %9
        \and 
        Research School of Astronomy \& Astrophysics, Australian National University, Canberra, ACT 2611, Australia} %10

\abstract
{
    NOTT (formerly Hi-5) is a new high-contrast L' band (3.5-4.0 \textmu m) beam combiner for the VLTI with the ambitious goal to be sensitive to young giant exoplanets down to 5 mas separation around nearby stars. The performance of nulling interferometers in these wavelengths is affected both by fundamental noise from the background and by the contributions of instrumental noises. This motivates the development of end-to-end simulations to optimize these instruments.
}
{
    To enable the performance evaluation and inform the design of such instruments on the current and future infrastructures, taking into account the different sources of noise, and their correlation.
}
{
    SCIFYsim is an end-to-end simulator for single mode filtered beam combiners, with an emphasis on nulling interferometers. It is used to compute a covariance matrix of the errors. Statistical detection tests based on likelihood ratios are then used to compute compound detection limits for the instrument.
}
{
    With the current assumptions on the performance of the wavefront correction systems, the errors are dominated by correlated instrumental errors down to stars of magnitude 6-7 in the L band, beyond which thermal background from the telescopes and relay system becomes dominant. 
}
{
    SCIFYsim is suited to anticipate some of the challenges of design, tuning, operation and signal processing for integrated optics beam combiners. The detection limits found for this early version of NOTT simulation with the unit telescopes are compatible with detections at contrasts up to $10^5$ in the L band at separations of 5 to 80 mas around bright stars.
}

\maketitle
\section{Introduction}
    \subsection{Nulling interferometry and the NOTT instrument}
        While thousands of exoplanets have been detected by indirect means, the direct detection of their emitted or reflected light is so difficult that only a few handfuls of giant, hot young planets on very wide orbits have been detected so far. This is in part due to the fact that the limited aperture of even optically perfect instruments produces an extended Point Spread Function as the image of point-like sources. As a result, the images of faint exoplanets are buried into the wings of the image of their host stars, and the associated photon noise dominates the signal to noise ratio (S/N). In the context of high-contrast detection, the resolving power can no-longer be expressed with simply a minimum angle, but must be considered as a map of the attainable contrast as a function of the position of the planet relative to the star.\par
        
        Coronagraphs have been developed and used to optically alter the response of instruments, effectively redirecting the on-axis light out of the science imaging channel. Such devices have reached a maturity in that design approaches and manufacturing tools are able to produce coronagraphs with inner working angles down to $0.5 \lambda/D$ of the collecting aperture \citep{Guyon2006a}.\par
        
        Interferometry offers a way to increase the effective extent of collecting apertures, therefore increasing the maximum angular resolution achievable. However, the measurement of the interferometric fringes, much like the recording of images, is plagued by the same effect of photon noise carried by the (virtual) PSF of the considered aperture array. In the case of long baseline interferometry, this PSF evolves with the projection of the array on the plane orthogonal to the line of sight. Compared to the PSF produced by a monolithic telescope, the sparsity of this array results in a finer bright pattern with wings extending over the full effective field of view.\par
        
        While pupil densification has been proposed to improve part of this slow falloff effect \citep{Tcherniavski2011, Guyon2005a}, recent works have shown \citep{Lacour2019,Nowak2020,Nowak2020a,GRAVITYCollaboration2021} that interferometry can be combined to the efforts of optically discriminating the off-axis light from the starlight  in the single-dish beams. In observations with the GRAVITY instrument \citep{Lacour2019}, the spatial filtering by single mode optical fibers used in each of the beams is used to isolate the light of a known exoplanet from most of the light of its host, before sending it to the beam combiner. On the detector, the planet light and the residual starlight form two distinct sets of fringes that can be effectively discriminated to reconstruct the spectrum of the planet.\par
        
        Yet, to take advantage of the longer baselines as coronagraphs take advantage of larger diameter apertures, a different approach to beam combination must be taken that gives the on-axis light a special treatment, so as not to let the photon noise it carries affect the scientific measurement. This approach is called nulling interferometry \citep{Bracewell1978}. In this approach, the phases of the combined beams are tuned to send on the main detector the dark fringe of the on-axis light, while the bright fringe is measured separately, as they are sent to different detectors. On the dark channel, called the interferometric null, light from an off-axis source may still get recorded, depending on its location in the field. \par
        
        This type of beam combination is difficult to use, since the path length in the different beams must be matched to within a fraction of the wavelength. Now that the interferometric facilities are equipped with high performance fringe tracking systems, the use of a nulling beam combiner becomes appealing \citep{Lacour2019a, Anugu2020a, Pannetier2021a}.\par
        
        The NOTT\footnote{Nott is the personification of night and darkness in Norse mythology. NOTT is also an acronym for Nulling Observations of exoplaneTs and dusT.} (formerly Hi-5) instrument is a proposed high-contrast visitor instrument for the VLTI \citep{Defrere2018, Defrere2018b, Defrere2022a}, that will make use of nulling beam-combination to allow the characterization of young giant planets in the L' band down to the snow line of their host system. It will leverage integrated photonic technology for beam-combination.\par
        
        Such photonic devices can implement optical waveguides integrated into monolithic glass substrates. Interference between the light from different apertures is obtained by coupling the waveguides, for example by bringing them close together so their evanescent field can be coupled \citep{Tepper2017}, or by fusing them into sections that guide a larger number of modes \citep{Soldano1995}. These different approaches offer different characteristics, in particular with respect to their chromatic behavior. Such devices can be extremely compact, which offers the possibility to build complicated combiner architectures in small and stable packages. \par
        NOTT is designed to operate as part of the Asgard suite \citep{Martinod2022}, a VLTI visitor instrument which also includes the HEIMDALLR high precision fringe tracker \citep{Ireland2018} and the Baldr wavefront sensor (based on Zernike wavefront sensor) which are currently under development.
        
    \subsection{Beam combination architecture}
        
        With the capability to leverage baselines of more than 100 m, other limitations have been identified. Indeed, as the stellar disk looses its spatial coherence for the baseline (i.e. as the stellar disk becomes resolved), some of this light leaks to the dark outputs, leading to confusion with the planet light. This has been an important driver for the evolution of nulling combiners from the original idea of \cite{Bracewell1978}, to involve more than two apertures with the Angel \& Woolf combiner \citep{Angel1997, Velusamy2003}, then even more complicated designs \citep{Guyon2013}. Yet, with the constraints of observing from the ground, the measurement errors due to varying optical path differences (OPD) are a dominating factor. More recent approaches have focused on this problem like by producing closure phase from the leakage light of a nuller \citep{Lacour2014}, or designing specific combinations that cancel out the second derivative of the phase errors with \cite{Martinache2018} in kernel nulling. In the light of the analysis proposed by \cite{Laugier2020d}, one can argue that the desirable robustness properties of kernel nullers apply to the sin-chop configurations of the Angel \& Woolf combiners. Indeed the two nulled outputs are constructed with contributions of all four inputs multiplied by vectors of phasors that are enantiomorph (of mirrored shape) of one another, like the pairs of outputs of kernel nullers or the outputs of complementary phase-chopping states investigated by \cite{Lay2004} and \cite{Defrere2010}. This gives these two output intensities correlated responses to instrumental errors, and gives their difference a desirable robustness to instrumental errors.\par
        
\begin{figure}
            \centering
            \includegraphics[width=0.45\textwidth]{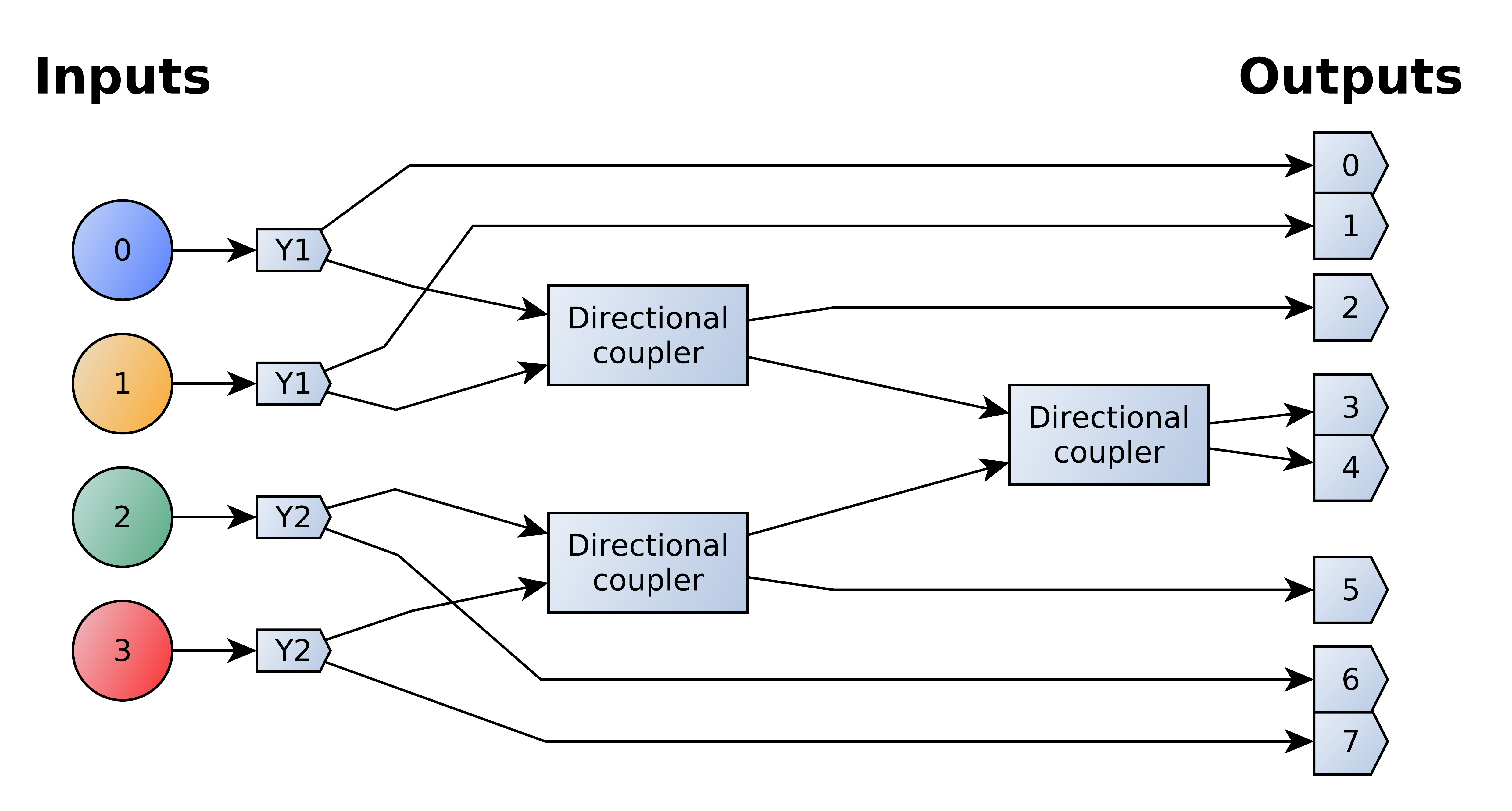}
            \caption{The schematic of the photonic beam combiner including, from left to right, the four inputs of the chip, and photometric taps labeled Y1 and Y2, the first combination stage composed of two directional couplers, and the second stage consisting of a single directional coupler mixing the dark outputs of the first stage. The chip has eight outputs, including photometric outputs (0, 1, 6 and 7), bright outputs (2 and 5) and two complementary dark outputs (3 and 4).}
            
            \label{fig:schema}
        \end{figure}

        Other advantages can be argued in favor of the simpler Angel \& Woolf design, like a higher raw throughput, and the possibility of external tuning, making it a technological stepping-stone towards the more complicated kernel-nullers that offer three times the number of observable quantities. The combination scheme anticipated for the photonic combiner of NOTT is illustrated in Fig. \ref{fig:schema} and includes photometric taps on each of the inputs for monitoring of the injection rate. \par
        
    \subsection{Treatment of the measurement errors}\label{sec:treatment}
        Our goal is to estimate the contrast and sensitivity performance of such a nulling combiner instrument to the presence of a planet around nearby stars. In the literature, this has been done in different ways.
        Some studies targeting space-based missions (e.g. \cite{Mugnier2006, Dannert2022}) focus exclusively on the fundamental noise sources, like photon noise, background noise and static leakage sources, and limit their conclusions to regimes that are dominated by these contributions.\par
        
        Important efforts were led by \cite{Lay2004} in the context of TPF-I to account for instrumental errors. They made use of second order expansion of the effect of the perturbations on the interferometric outputs and remain a reference for this type of study. However, this is limited to the regime of small aberrations where first and second order terms dominate, and to the case of ideal beam-combination with a continuously rotating array. In the case of a ground-based instrument working at smaller wavelengths the atmospheric contributions are expected to play a dominant role in overall errors as was highlighted in the case of GENIE \citep{Absil2006}, and ALADDIN \citep{Absil2007}. Furthermore, the instrumental noise is expected to be highly correlated between different wavelength bins, an effect that has in large part been neglected.\par
        
        Works by \cite{Ireland2013a}, \cite{Ceau2019b} and \cite{Kammerer2020a} have highlighted the performance benefits of accounting for correlations in interferometric data. In \cite{Thiebaut2006, Mugnier2006, Dannert2022}, only the fundamental noise sources (such as photon noise) are considered. Those are independent in classical spectrographs, as the spectral channels are recorded in different pixels. \cite{Lay2004} mentions the temporal correlations between spectral channels but does not take them into account in the determination of the overall performance. The straightforward performance analysis offered by \cite{Martinache2018} considers broadband observations (a single spectral channel), and does not account for correlations that would arise between the three kernel-null outputs. To our knowledge, the correlations in the nulling interferometry data have only been considered by \cite{Lay2006} in an approach that relies on this correlation to propose an ad-hoc approach of polynomial fitting and subtraction. The author then concludes that the success of this approach depends on the geometry of the array and the resulting temporal behavior of the signal of interest.\par
        
        To enable the exploration of the opportunities offered by advanced combiners, we developped a new simulation tool, specifically tailored to complicated photonic beam combiners, and compatible with the demands of nulling. Here, we describe SCIFYsim\footnote{\url{https://github.com/rlaugier/SCIFYsim}}, a new end-to-end simulator package written in python and including Fourier optics tools for its bulk optics components and complex amplitude matrix manipulation for its single-mode components. We then show some of the early results that it provides in the effort to evaluate the performance of NOTT.\par
        
        Nulling self calibration \citep{Hanot2011, Defrere2016, Mennesson2016, Norris2019, Martinod2021} is a different approach making use of the empirical distribution of the interferometric outputs to produce a statistical inference. To our knowledge, it has never been deployed for nullers more complex than the Bracewell nullers. Considering its implications in the context of NOTT is outside the scope of this work.\par

        In Sect. \ref{sec:combination} we describe the computation produced by SCIFYsim, before listing in Sect. \ref{sec:aberrations} the aberration affecting the input beams. We then examine the distribution of the differential nulled output in Sect.  \ref{sec:errors} and describe statistical tests for their interpretation in Sect. \ref{sec:tests_new}. In Sect. \ref{sec:application}, we use these tests to evaluate the performance of the NOTT instrument at VLTI for different stars and different configurations, before proposing a discussion and conclusion in Sects. \ref{sec:discussion} and \ref{sec:conclusion}.
        
\section{The SCIFYsim end-to-end simulator}\label{sec:combination}
    \subsection{The architecture of the simulator}
        This gives a short overview of the architecture of the simulator. More details on some of its components are given in the following sections. The simulator articulates around a number of modules and classes that are combined and used by a \verb+director+ class, in order to compute and hold the parameters of a model observatory, instrument and observing session. Here is a partial list:
        \begin{itemize}
            \item \verb+observatory+ holds the geometric configuration of the array, and relies on the astroplan and astropy packages to compute the location of the target in the sky at a given time, the resuling alt-azimuthal pointing, and the projection of the array along --- and orthogonal to --- the line of sight (see Sect. \ref{sec:array_geom_source}).
            \item \verb+sources+ is used to compute spectra of absorption and emission of the targets, and the parasitic contribution of the earth atmosphere and optical train (see Sect. \ref{sec:throughput_bg}).
            \item \verb+correctors+ contains tools for delay lines and longitudinal dispersion compensators (see Sects. \ref{sec:dry_OPD} and \ref{sec:chromatic_opd}).
            \item \verb+combiners+ is used to build and hold models of the beam combiner integrated optics (see Sect. \ref{sec:combiner_models})
            \item \verb+spectrograph+ contains classes to reproduce the behavior of the spectrograph and detector pixels (with the \verb+integrator+ class), with associated quantum efficiency, dark current, readout noise (see Sect. \ref{sec:spectrograph_detector}).
            \item \verb+analysis+ contains tools to handle the data reduction, and compute the performance of the instrument (see Sects. \ref{sec:errors} and \ref{sec:results}).
        \end{itemize}

        Each astrophysical source starts as vector of real valued spectrum of emission collected by each telescope. In accordance to its position in the field of view it is associated with a complex phasor encoding for each telescope:
        \begin{itemize}
            \item The phase effect due to its geometric position away from the optical axis of the array,
            \item the phase and amplitude errors due to fluctuating optical aberrations (Sects. \ref{sec:dry_OPD} and \ref{sec:chromatic_opd}),
            \item the amplitude effect due to the near Gaussian transmission map of the spatial filtering by the single mode waveguide, limiting the field of view,
            \item the amplitude effect due to optical transmission of the sky and optics.
        \end{itemize}

        The complex amplitude of each source is multiplied by this phasor to obtain the complex amplitude of the light entering the integrated beam combiner. This is then multiplied by the matrix of the beam-combiner (see Sect. \ref{sec:combiner_models}) to obtain the complex amplitude at the outputs. This is computed for each wavelength bin, and for subexposures of a few ms which are summed over the detector integration time. Values can be retrieved here under different forms, for example as a total sum including dark currents, photon and readout noise as in a realistic observation, or as exact values for each contributing sources in order to compare useful signals from error terms and evaluate performance (see Sects. \ref{sec:errors} and \ref{sec:tests_new}).\par

        Due to their diffuse and incoherent nature, the parasitic thermal background sources are not computed at each time step, but are instead updated only when necessary, to save computing time.
        
    \subsection{Chromatic combiner models}\label{sec:combiner_models}
        The flexibility of SCIFYsim comes in part from the use of complex amplitude transfer matrix to model the beam combiner. Such matrices have been used to model the most complex beam combiner architectures, especially in the context of nulling \citep{Guyon2013, Martinache2018, Laugier2020d}, and are well suited for modeling the interference of single mode beams. They take as input a vector containing the complex amplitude of the incoming beams, and produce the vector of complex amplitudes at the outputs.\par
        
        In SCIFYsim, they can be built in different ways. Either based on \texttt{sympy} symbolic expressions such as were used in \citet{Laugier2020d} and the corresponding assortment of combinatorial, algebraic and analytical tools\footnote{\url{https://github.com/rlaugier/kernuller}}. Alternatively they can be assembled algebraically based on building-block couplers of Fig. \ref{fig:schema}, each represented by a matrix of the form:
        \begin{equation}
            \mathbf{M}_{\mathrm{coupler}} = \left[\begin{matrix}\sqrt{b} & - \sqrt{1 - b} e^{- j \Delta}\\\sqrt{1 - b} e^{j \Delta} & \sqrt{b}\end{matrix}\right],
        \end{equation}
        where $b$ is the coupling rate, and $\Delta$ is the incurring phase offset, both wavelength dependent.\par
        
\begin{figure}
            \centering
            \includegraphics[width=0.45\textwidth]{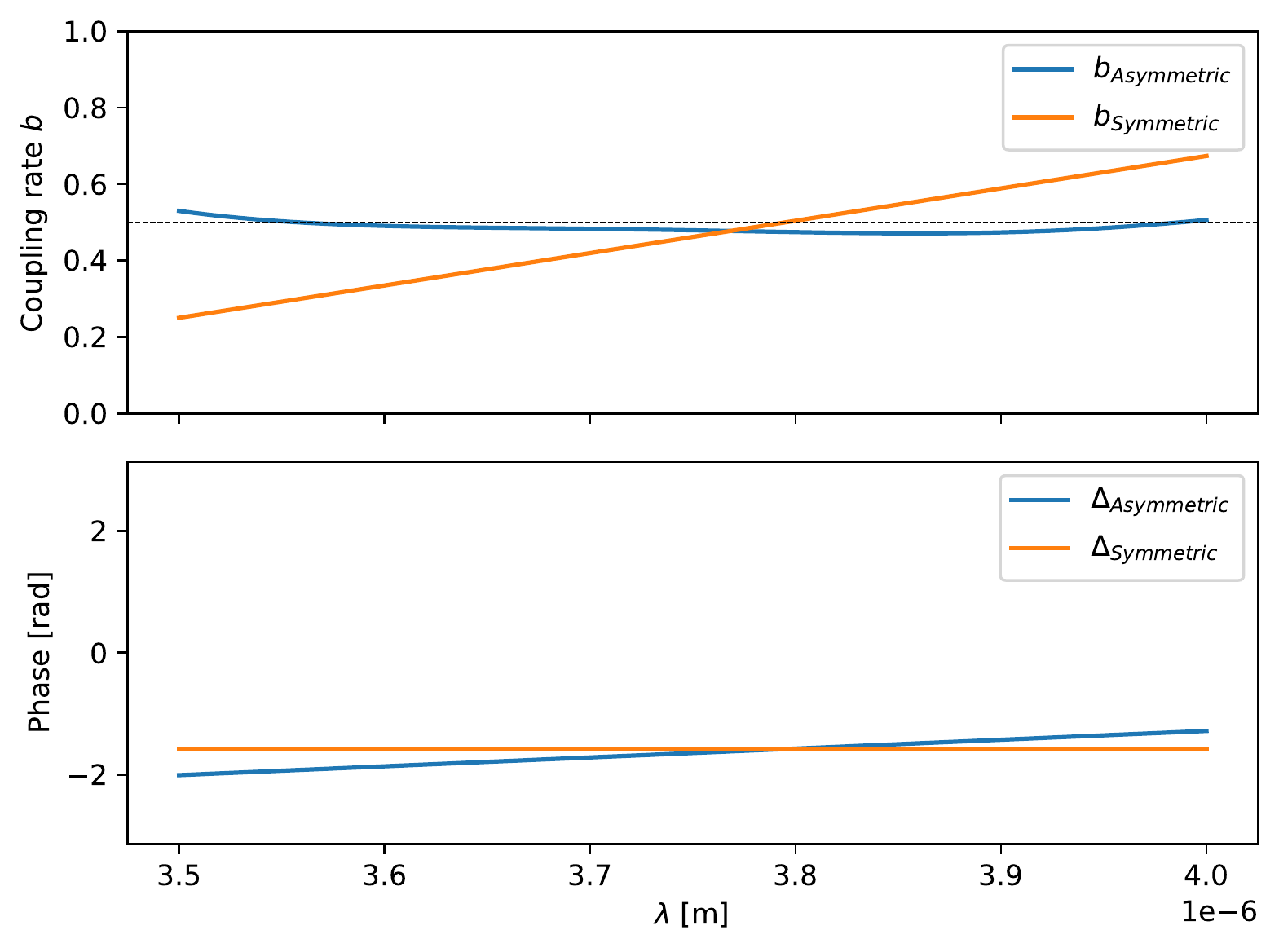}
            \caption{The behavior of the asymmetric combiner used as building-block for the instrument in comparison to the symmetric alternative. Asymmetry in the evanescent wave coupler to produce the flatter coupling rate at the cost of a chromatic phase offset.}
            \label{fig:combiner_plot}
        \end{figure}

        This informs the model on its internal architecture, and by replacing the matrix of the building-block coupler by a chromatic model (i.e. a parametric matrix as a function of the wavelength $\lambda$) informed by the characterization of one or other combiner technology \citep{Tepper2017, Sharma2020}, one obtains the chromatic model of the combiner as a whole. The model used here was prepared with beam propagation simulations and experiments made at Macquarie University for the relevant technology, and its behavior is illustrated in Fig. \ref{fig:combiner_plot}. The matrix is then evaluated for the relevant wavelength channel and stored in a 3D array of complex float for use in the numerical computations.\par
        
\begin{figure}
            \centering
            \includegraphics[width=0.45\textwidth]{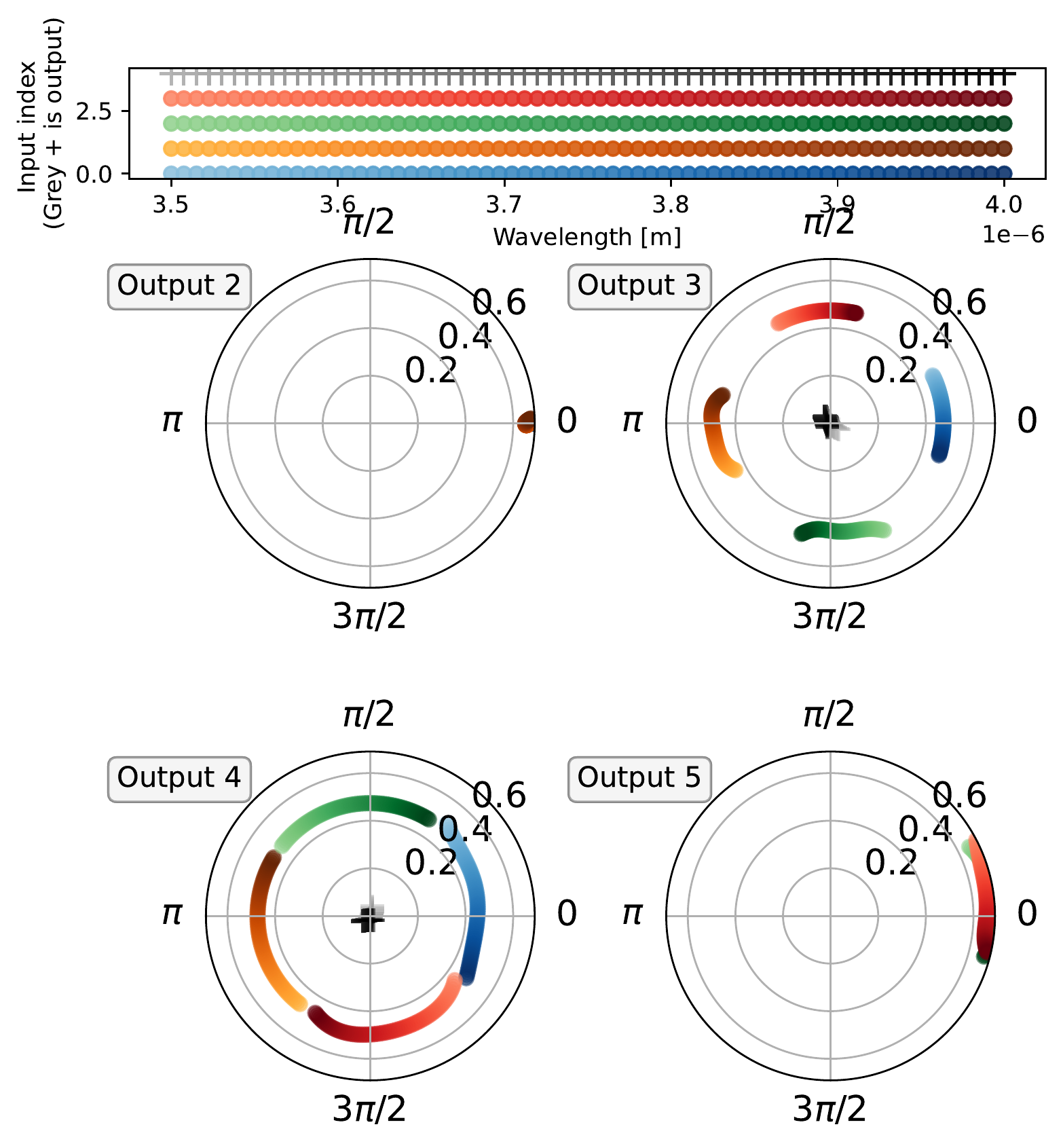}
            \caption{The complex matrix plot of the matrix after correction. Each plot represent the complex plane for one output, each input is represented by a hue, and the luminance relates to the wavelength. A gray "+" marks the complex sum of the contribution, giving the nominal complex amplitude of the outputs. For the bright outputs, the colored spots land on top of one-another, and the output "+" falls outside of the range.}
            \label{fig:cmp_corrected}
        \end{figure}

        In the case of fringe measuring instruments like PIONIER \citep{Berger2010} and GRAVITY \citep{GravityCollaboration2017}, with their outputs dispersed by a spectrograph, the visibility can be computed independently in each wavelength so that intensity or phase imbalance arising from chromatic aberrations have no first order impact on performance. In the case of nulling interferometry, the rejection of the photon noise happens from a precise control of the phase of the interference, which must be set throughout the bandwidth of the instrument. In order to represent and examine the resulting complex null configuration, we extend the representation used in \citet{Laugier2020d} to make use of color maps to represent simultaneously all the wavelength channels of the instrument, as can be seen in Fig. \ref{fig:cmp_corrected}. Here, arrows are replaced by colored spots, with a shading indicating the different wavelengths. On these same plots, we represent the sum of the contributions as an additional spot in shades of gray for each wavelength. Gray spots more tightly packed around the origin represent deeper an more broadband nulls.\par
        
    \subsection{Spectrograph and detector}\label{sec:spectrograph_detector}
        The spectrograph module takes the intensity computed at the outputs of the combiner chip and convolves it with a spectroscopic point spread function (PSF) that spreads the light, depending on its wavelength on the 2D array of pixels of the detector. To accommodate the possibility of undersampled spectroscopic focal plane, this is first done with a denser array, then binned to the final detector resolution.\par

        While helpful for the validation of the spectrograph design and the validation of signal extraction methods \citep{Dandumont2022}, this is demanding in computation resources. For the needs of large Monte-Carlo samples, this step is bypassed, relying instead on the assumption that each spectral channel is spread uniformly over a fixed number of pixels.\par
        
        The integrator module simulates the behavior of the detector pixels. It adds the photon noise, readout noise and dark current. It can also implement saturation effects, or amplification and the associated excess noise factor such as will occur in the case of EMCCDs or avalanche photodiodes detectors \citep{Lanthermann2019}. For NOTT, it implements the properties of the Teledyne HAWAII2RG detector with $15~e^-$ of readout noise and $0.02 ~e^-s^{-1}\mathrm{pix}^{-1}$ of dark current.\par
        
        Recordings of the OPD, or other aberrations can also be saved to this object as metrology outputs.
        
    \subsection{Throughput and background signal}\label{sec:throughput_bg}
        For flexibility, the throughput and thermal background signal are both evaluated based on the series of objects called transmission-emission objects. Each is affected a temperature and a transmission function (either flat, or tabulated), in a way similar to what was done in GENIEsim \citep{Absil2006} for a two-telescope Bracewell architecture. These include:
        \begin{itemize}
            \item the sky,
            \item the UT or AT optics (including the delay line train),
            \item the warm optics of the instrument,
            \item the combiner chip,
            \item the cold optics (including the spectrograph).
        \end{itemize}
        Here, they are linked into a chain by pointers so that they each provide methods to obtain, recursively, values like the total throughput of the chain downstream from them, with commands such as \texttt{item.get\_downstream\_transmission()}.\par

        \begin{table}[]
            \small
            \centering
            \begin{tabular}{c c c c}
                \hline
                \hline
                
                Source & Temperature & Mean & Contribution\tablefootmark{a}\\ 
                 & [K] & transmission & $[e^- s^{-1}]$\\
                
            \hline
            Sky & 285.0 & 0.92 & 1.20e+05 \\
            UT optics & 290.0 & 0.31 & 2.91e+06 \\
            Warm optics & 285.0 & 0.60 & 2.20e+06 \\
            Combiner & 100.0 & 0.50 & 1.69e-04 \\
            Cold optics & 100.0 & 0.56 & 2.65e-04 \\
            Enclosure & 100.0 & N.A. & 4.23e+05 \\
            Dark current & 60.0 & N.A. & 1.07e+01 \\
            \hline

            \end{tabular}
            \caption{The components of the radiometric chain given for the medium spectral resolution R=400.}
            \label{tab:radiometry}
            \tablefoot{
            \tablefoottext{a}{On one dark output, cumulated over all the wavelength band.}
            }
        \end{table}
        
        Like in GENIEsim, the thermal background is computed using the Planck law and an emissivity value taken as $1-t$ where $t$ is the functional transmission (reflectivity of mirrors and transmission of lenses). It assumes that the single mode filtering is done right at the end of the chain so that the étendue obtained in each spectral channel corresponds to the étendue injected in the waveguides. This is an approximation that relies on the way the pupil of the spectrograph samples the Gaussian beam that propagates from the outputs of the combiner.\par
        
        The exception to this is the enclosure term corresponding to the cryostat, which is purely radiometric and computes a flux of photons per detector surface, and therefore per pixel. The list of contributors is given in Table \ref{tab:radiometry}, with the total resulting current given for the sum of all spectral channels for R=400, including the dark current. The total optical throughput is about 17\% from the input of the Asgard table, or 5.2\% from the VLTI telescopes, leading to an overall efficiency of about 3.2\%, including the quantum efficiency of the detector. The total contribution of shot noise increases with the spectral resolution as the number of spectral bins increases. 
        
    \subsection{Array geometry and source}\label{sec:array_geom_source}
        The horizontal and projected geometry of the array is handled by a separate module called \texttt{observatory}. This is powered by \texttt{astropy} and \texttt{astroplan} to handle the coordinates of targets, pointings, and time. It also provides convenience functions to locate targets by name, or identify the best night of the year to make the observations i.e. for which transit of the meridian happens in the middle of the night). The pointing of the array also informs an airmass parameter inside the sky object.\par
        
        This module computes the projection of the array onto the plane orthogonal to the line of sight, which is then used to compute the geometric OPD affecting the beams of each light source depending on its position $(\alpha, \beta)$ in the field of view.
        \begin{equation}
            \tilde{a}_l = a_{l} e^{\frac{2 j \pi \left(X_{l} \alpha + Y_{l} \beta\right)}{\lambda}}
        \end{equation}
        where for each aperture $l$, $a_{l}$ is the amplitude collected, $X_l$ and $Y_l$ are the coordinates of the apertures projected in a plane orthogonal to the line of sight and aligned to right ascension and declination. In the example presented here, the star is modeled as a dense collection of points on a disk, with a black body spectrum.\par
        
        The module also computes the projection of the array onto the line of sight, therefore providing the relative path length to be corrected by the delay lines.
        
\section{Description of the main sources of errors}\label{sec:aberrations}
    \subsection{Injection into single mode fibers}
        The injection into single mode waveguides poses a challenge for any single-mode instrument. Indeed, aberrations in the beam represented by small amplitude or phase errors across the pupil produce small variations in the focal plane complex amplitude, which will result in decreased coupling into the mode field of the waveguide. These errors are the result of pointing errors and residuals from the correction of the atmospheric turbulence by the adaptive optics.\par
        
        As illustrated in Fig. \ref{fig:schema}, the amount of light injected into the combiner is monitored for each input with an evanescent coupler sending 10\% of the guided light to the photometric outputs. These enable the tuning of the injection and its monitoring in real time. While they will be invaluable in the field for flux calibration, their output is not used in the signal processing described here.
        
    \subsection{Residual OPD errors from fringe tracking}\label{sec:dry_OPD}
        The optical path errors are introduced as piston to each of the input beams. Random time series of such piston are computed based on the power spectrum of measurements obtained with the fringe tracker of GRAVITY on a bright star. For this, we take the square root of this power spectrum and associate it with random phase following a uniform distribution between $-\pi$ and $\pi$ then put it through an inverse Fourier Transform to obtain a random time series of the same power spectrum.\par
        
        The amplitude of this residual on the Unit Telescopes (UT) have been identified to exceed the expected values produced by the filtered atmosphere \citep{Lacour2019a}. This was attributed to vibrations in the telescope structure. An ongoing effort as through SCIFY as part of the GRAVITY+ upgrade is expected to bring this residual down to 100nm RMS of optical path difference, which corresponds to the level already reached on the Auxiliary Telescopes (AT). For this reason, we compute this optical path temporal power spectrum from observations made with the ATs, and use them in simulations using the UTs.\par
        
    \subsection{Atmospheric dispersion and chromatic OPD}\label{sec:chromatic_opd}
        In addition to the "dry" (i.e. purely geometric) piston described in Sect. \ref{sec:dry_OPD}, the phase of the inputs is also affected by a chromatic or "wet" contributions. This is important as the delay lines compensate, in ambient air, a path difference that occurs in the vacuum of space, and results in longitudinal atmospheric dispersion. This chromatic effect is modeled in the same way as GENIEsim, based on the tables by \cite{Mathar2004}. This static effect was not included in the early results presented here. NOTT will include ZnSe variable thickness corrector plates \citep{Koresko2006} to compensate this effect to the first order.\par
        
        While the contribution of turbulent air is often only associated to simple geometric piston as described in Sect. \ref{sec:dry_OPD}, the demands of a broadband nulling requires the correction for a more thorough model like described by \cite{Colavita2004} and \cite{Absil2006}. However, relevant simulations of these effects will depend on the technical solutions retained for HEIMDALLR and Baldr modules of Asgard. \cite{Colavita2004} predicted the effect of water vapor seeing for Mauna Kea to be 1/20th of the dry effect in the near infrared and  1/7th at $10~\mathrm{\mu m}$. In this work, dynamic chromatic perturbations are neglected but they will be included in later work, based on expertise gathered from GRAV4MAT at VLTI and METIS for the ELT.% <> Angel: most humidity is in the first km of atmosphere... <> More recent examination by Absil.
        
    \subsection{Tuning internal combiner chromatic errors}\label{sec:tuning}

        Chromatic biases in the combiner are modeled with polynomial fits of order six made from beam-propagation simulations and experimental results making available symbolic expressions of the beam-combiner matrix, and enabling flexibility in the choice of spectral channels. Options include some baseline implementations of symmetric and asymmetric evanescent wave couplers and multimode interference couplers. The result is an imperfect nulling combination due to chromatic phase and amplitude variations. As shown in Fig. \ref{fig:cmp_corrected}, a common phase term affects the whole of the dark outputs, but is irrelevant to our measurement of intensity. For each wavelength, the orthogonal four-way pattern is respected, and close to symmetrical.\par
        
        The variable thickness plates mentionned in \ref{sec:chromatic_opd} are used to compensate the chromatic errors in the combiner. The model used is the same as that used in GENIEsim \citep{Absil2006} and reproduces a vector $\mathbf{c}$ of complex phasors 
        \begin{equation}
            \tilde{c}_l(\lambda) = c_l e^{\frac{2 j \pi }{\lambda} \left(n_{\mathrm{air}}(\lambda) p_l + n_{\mathrm{glass}}(\lambda) q_l\right)}
        \end{equation}
        where $n_{\mathrm{air}}$ and $n_{\mathrm{glass}}$ are the refractive index of air and glass, $p_l$ and $q_l$ are the additional optical path in air and glass respectively. The term $c_l$ can be used to represent filters or adjustable stops, but are left to 1 in this study. This leads to an effective combiner matrix that can be expressed as : 
        \begin{equation}
            \mathbf{M} = \mathbf{M_0}.\mathrm{diag}(\mathbf{c})
        \end{equation}
        which includes the effect of the phasor $\mathbf{c}$ on each of the inputs of the combiner, distinguished from  $\mathbf{M_0}$, the matrix representing the static part of the beam combiner.\par
        
        The air and glass path lengths are both adjusted by gradient descent in two independent steps involving different subsets of beams and different cost functions. The first can be interpreted as maximizing the null depth obtained by the first nulling stage by the relative phase between 0 and 1 and between 2 and 3. Using a gradient descent method, the ZnSe thickness and geometric path length are adjusted in order to minimize the cost vector $\mathbf{g}_{\mathrm{null}}$ based on the values of $I_k$, the intensity at output $k$:
        \begin{equation}
            \mathbf{g}_{\mathrm{null}} = \frac{\sum_{k \in \mathrm{dark}} I^-_k{}(\lambda) }{\sum_{k \in \mathrm{bright}} I^+_k(\lambda) },
        \end{equation}
        with one dimension per spectral channel. This is done using a Levenberg-Marquardt algorithm.\par
        
        The second step can be described as tuning the second (mixing) stage and can be done from from upstream, by acting differentially on the input pairs (0, 1) - (2, 3). To do so, we need a cost function that reflects the desired symmetry quality between the complementary combinations that would guaranty the robustness quality of the observable, as outlined in \cite{Laugier2020d}. For this, we compute a shape parameter:
        \begin{equation}
            \Lambda =\frac{\underset{\theta \in [-\pi, \pi]}{\mathrm{min}} \Big( ||e^{j\theta} \cdot \mathbf{m}_{h} - \mathbf{m}_{k}^* ||^2 \Big)} { \Big( \sum_{i=0}^{N} |m_{k,i}| \Big)^2}
        \end{equation}
        with one dimension per spectral channel per differential output, where $\mathbf{m}_h$ and $\mathbf{m}_{k}$ are a pair of rows of the corrected combination matrix $\mathbf{M}$ corresponding to a pair of complementary dark outputs. $\theta$ is a dummy variable representing the fact that only the output intensity is recorded and the consideration must therefore be blind to an arbitrary phase offset between the two rows. A normalization is made by the sum of the coefficient in one of the rows which represents the peak amplitude of the response map. Again, this is minimized numerically with a Levenberg-Marquardt algorithm by adjusting lengths in air and glass.\par
        
        In order to use this parameter in practice, the matrix of the combiner would have to be measured using the calibration source. This can be obtained through a process comparable to the one used to measure V2PM matrices \citep{Lacour2008, Cvetojevic2021} for ABCD combiners. Further work should make this process reliable and accurate.\par
        
\begin{figure}
            \centering
            \includegraphics[width=0.45\textwidth]{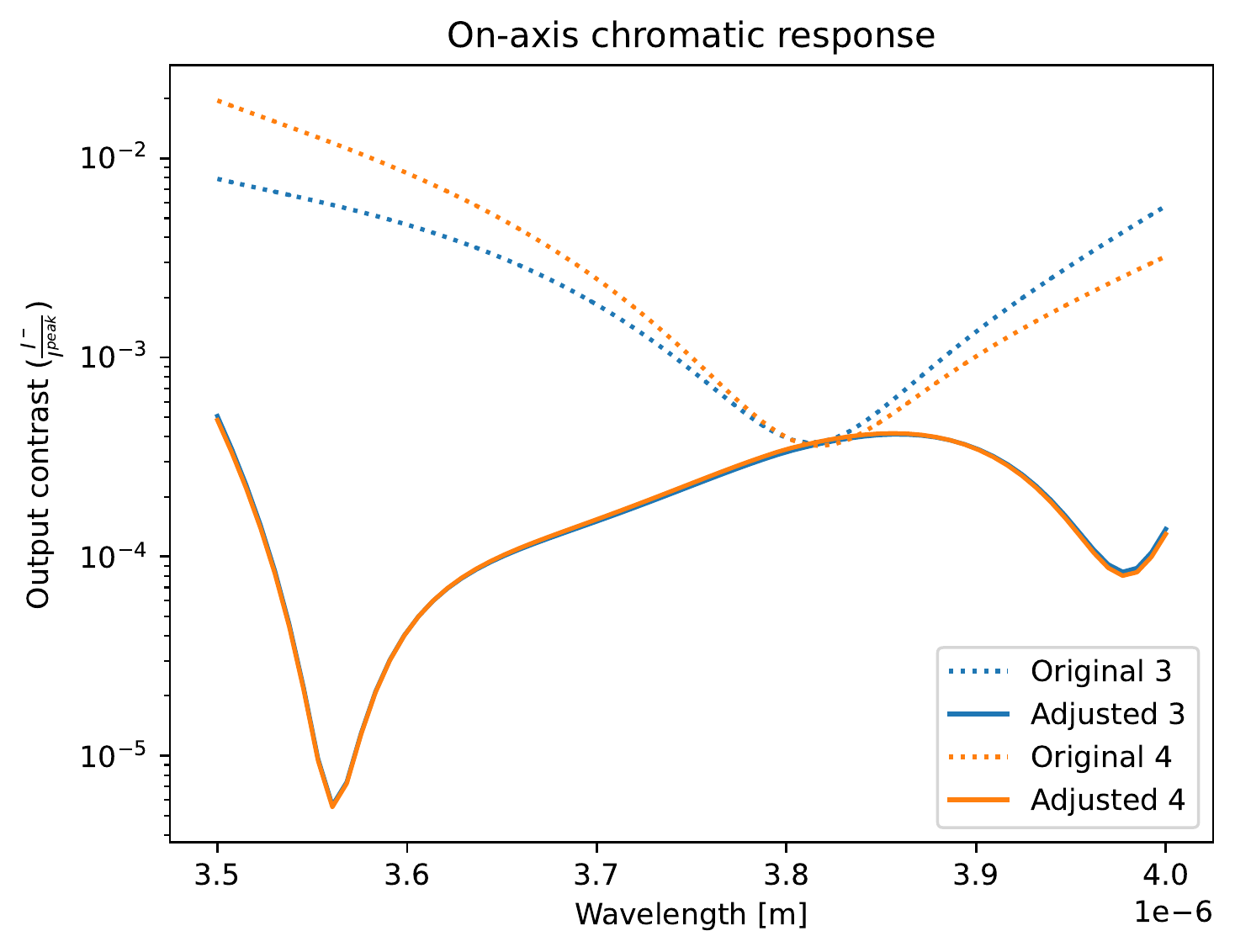}\par
            \includegraphics[width=0.45\textwidth]{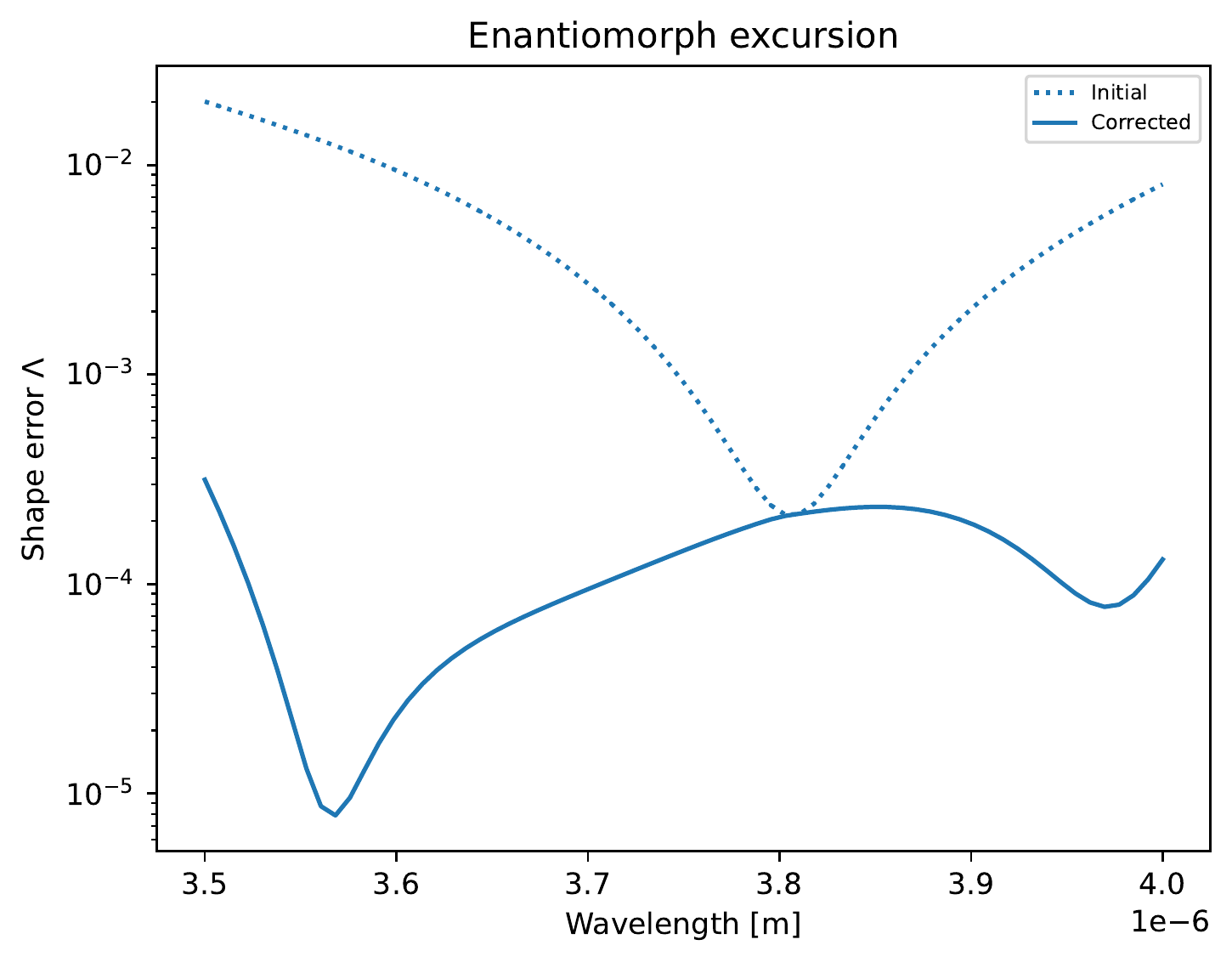}
            \caption{The cost function for the adjustment of the delay lines and dispersion corrector. The null depth of both dark outputs labeled 3 and 4, guiding the optimization for the first stage of the combiner (top) and shape error guiding the optimization for the second stage (bottom) before and after adjustment in dotted line and plain line respectively.}
            \label{fig:null_and_shape}
        \end{figure}

        The null depth and shape parameter for the initial and final state plotted in Fig. \ref{fig:null_and_shape} show the wavelength range split into three different regions by two troughs of null depth and shape parameter. These deeper minima in this broadened favorable region have an impact on the instrumental noise described in Sect. \ref{sec:parametric_errors}.
        
\section{Examination of the resulting errors}\label{sec:errors}
    \subsection{Time series of the raw combiner outputs}
\begin{figure}
            \centering
            \includegraphics[width=0.45\textwidth]{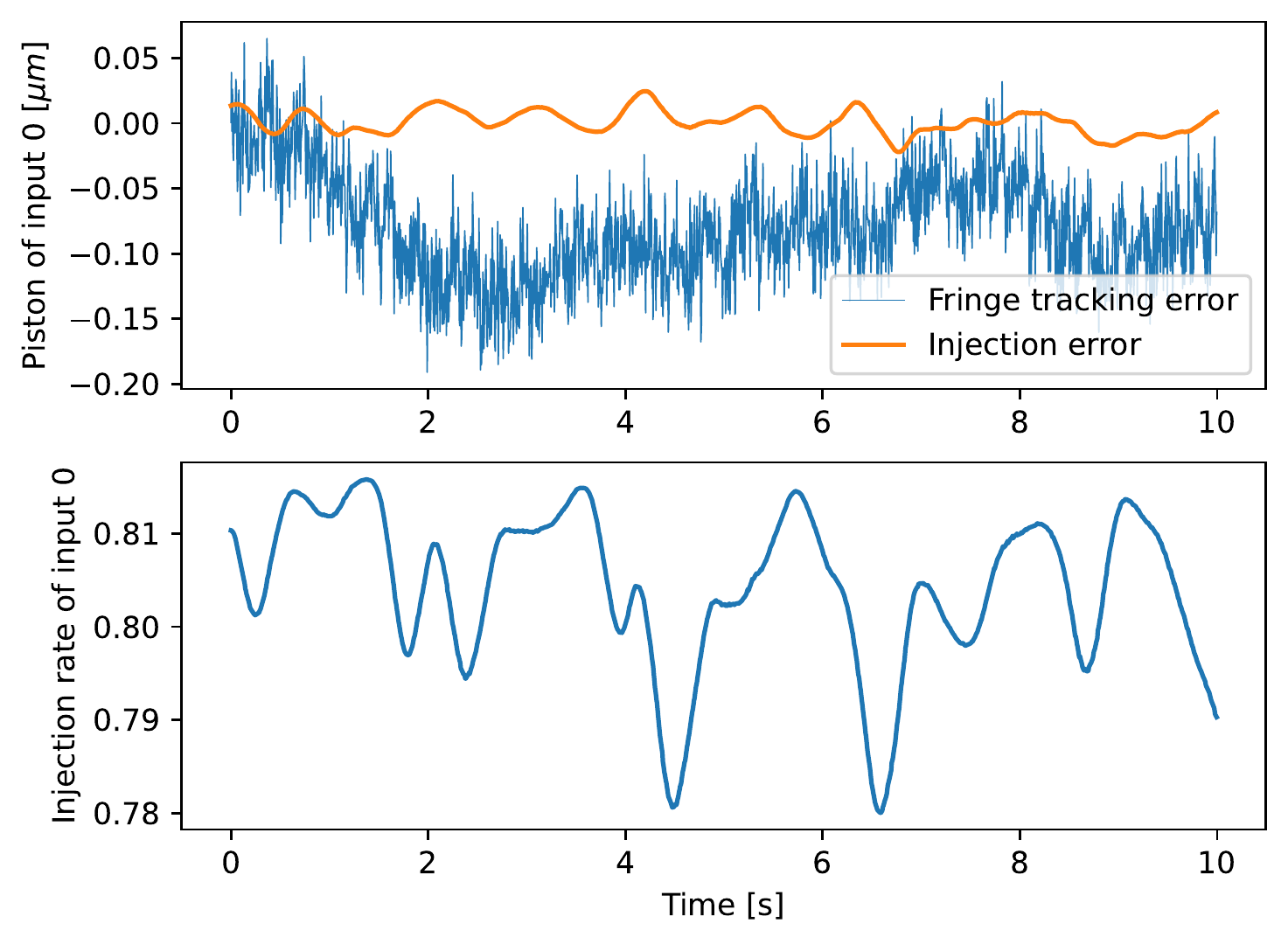}
            \caption{High temporal resolution series of the instrumental noise contribution on one input. Output 3 and 4 are the two nulled outputs. This shows the time resolution of the simulation which includes two different sources of phase errors: from fringe-tracking and from wavefront errors at the injection; as well as the injection rate error.}
            \label{fig:time_series_input}
        \end{figure}

        To illustrate the behavior of the simulated instrumental errors, we track the temporal variations of the contributions to input piston and amplitude. These are displayed in Fig. \ref{fig:time_series_input} for one of the four inputs. Note that, for this illustration, a single piston is considered rather than an optical path difference between the collectors of a baseline. The temporal spectrum for the simulated effects of injection contains little high frequency content as expected for averaging of large diameter apertures, but the temporal spectrum of the fringe tracking error contains significant high-frequency content, perhaps inflated by measurement noise since it was generated from real data.\par
        
\begin{figure}
            \centering
            \includegraphics[width=0.45\textwidth]{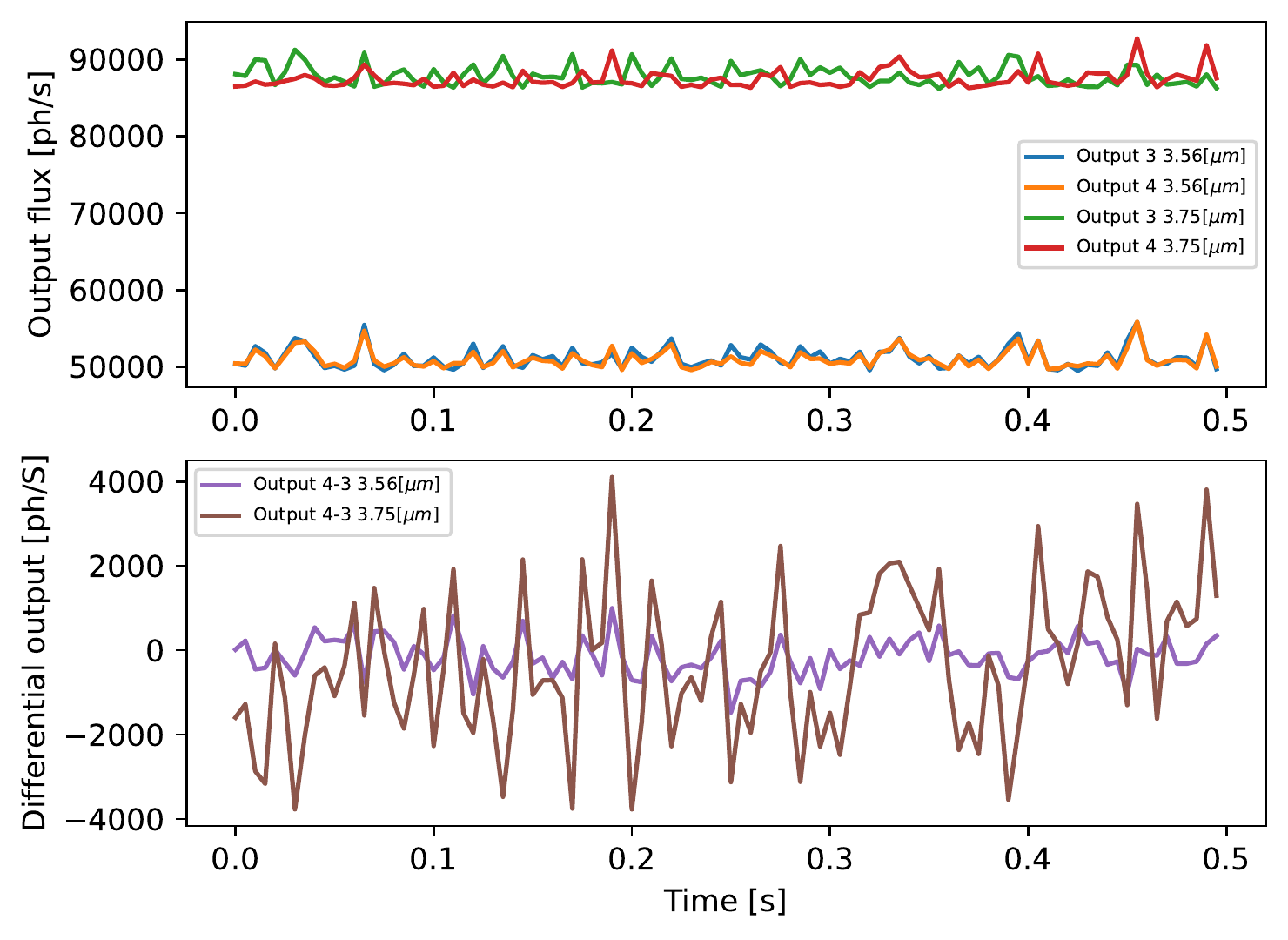}
            \caption{High resolution time-series of the dark outputs (top) and their difference (bottom) for two different wavelength bins offering different qualities of combination, leading to different mean leakage and correlation level. Correlation within each pair is the effect leading to self-calibration in the differential output. Correlation between spectral channels is what prompts for the data whitening approach. The time series shown here are indicative only since in practice integrations are made over longer times and are contaminated by sensor noises before they can be numerically subtracted.}
            \label{fig:time_series_output}
        \end{figure}

        The effects of these aberrations on the dark outputs are shown in Fig. \ref{fig:time_series_output} for two different spectral bins: $3.75~\mathrm{\mu m}$ at the center of the band, and $3.56~\mathrm{\mu m}$ down one of the troughs identified in Sect. \ref{sec:tuning}. This representation highlights the practical effect of the self-calibration of instrumental errors in the differential null. Correlation of the error terms is higher at $3.56~\mathrm{\mu m}$ as anticipated from the combination closer to its nominal scheme, leading to smaller errors in the differential output. Note that the values showed here are pure measurement of intensity on the detector simulated at short $5~\mathrm{m}s$ time steps. The effect of shot noise, dark current, read noise and longer integration time are not shown here but are included in the rest of this study.
        
    \subsection{The distribution of errors}
        As remarked by \citet{Martinache2018}, a significant advantage of the differential nulls over the simpler Bracewell nulls is their distribution. Indeed, the stochastic instrumental errors, dominated by the variation of the leakage light under varying optical path differences and amplitude fluctuations are symmetrical. When the errors are small, the distribution approximates to a Gaussian. SCIFYsim can be used to evaluate the limits of this approximation.\par
        
\begin{figure}
            \centering
            \includegraphics[width=0.45\textwidth]{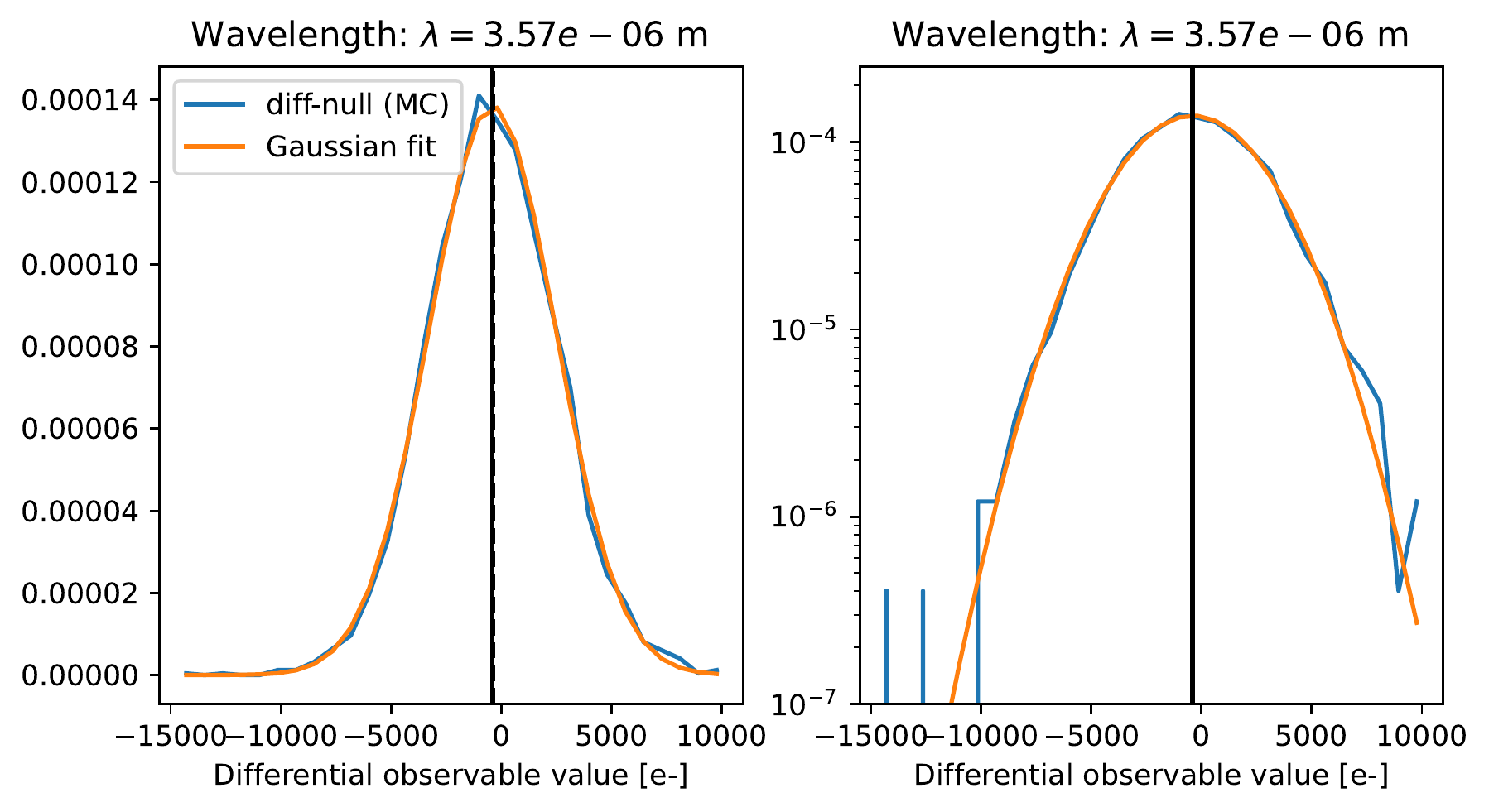}\par
            \includegraphics[width=0.45\textwidth]{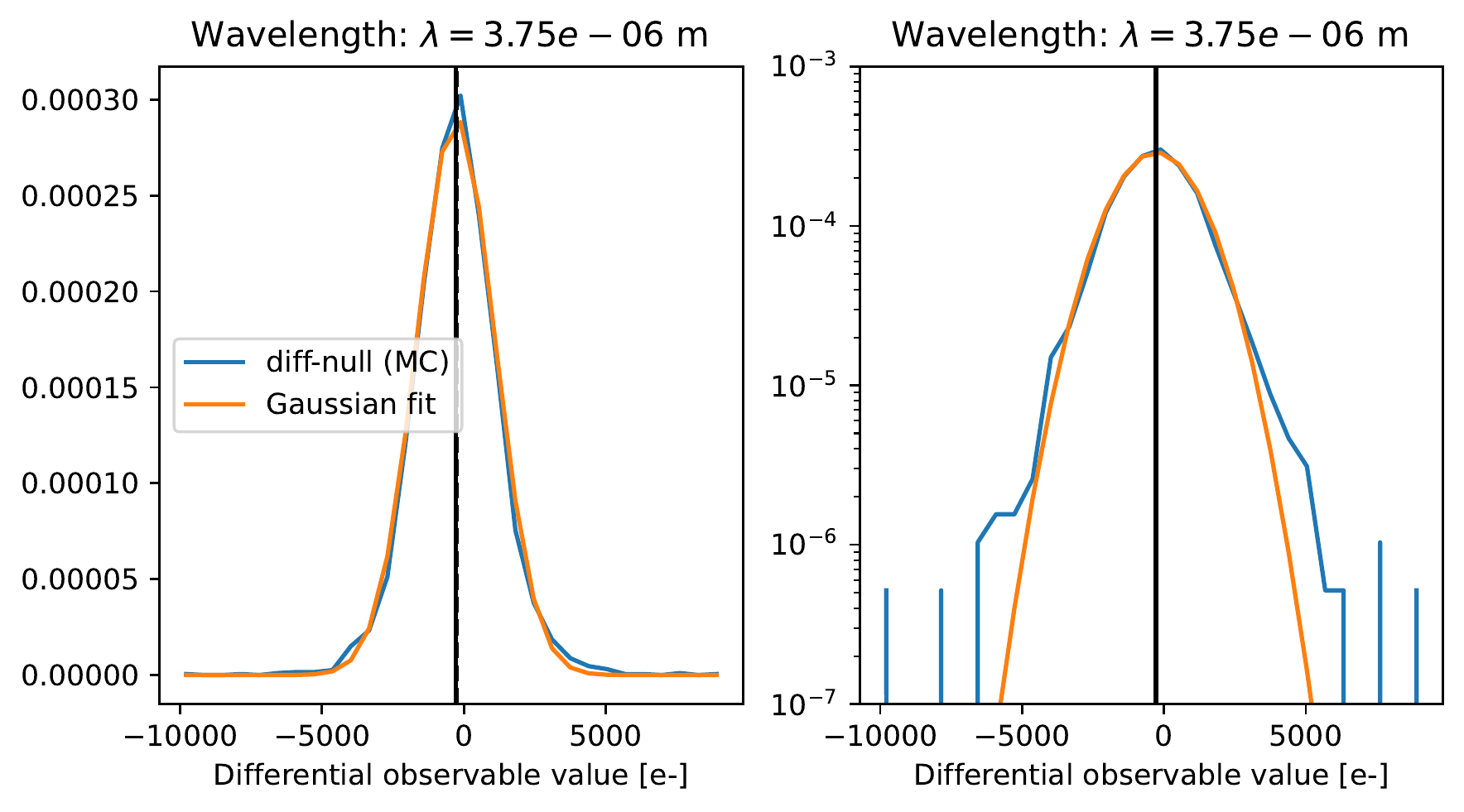}
            \caption{Top: Empirical probability density function of the total error for the 3.57 and 3.75 $\mathrm{\mu m}$  bin computed with 3,000 samples of 3 s dits with spectral resolution $R \approx 400$. It is displayed in linear scale (left) and in log scale (right). Some kurtosis is noticeable in log scale causing the Shapiro-Wilk test to reject the Gaussian hypothesis in parts of the spectrum but the Gaussian distribution remains a relevant approximation.} 
            \label{fig:distribution}
        \end{figure}

        We examined the distribution of the differential null, as illustrated in Fig. \ref{fig:distribution} for two different spectral bins. The quality of the distribution is quantified with the p-value of a Shapiro-Wilk test, giving the probability of wrongly rejecting the null hypothesis (Gaussian hypothesis). We consider that when this number drops below five percent, the Gaussian hypothesis should be rejected. For the parameters considered here, the pdf looks Gaussian in linear scale but the test starts to reject the Gaussian hypothesis with about 400 samples of the 3 s DIT for most spectral channels. However, the Gaussian distribution remains a good approximation, especially for the mean of a large number $n_{\mathrm{DIT}}$ of samples as used here, and will be used in the rest of this work, in particular the hypothesis tests detailed in Sect. \ref{sec:tests_new}.\par 
        
    \subsection{Parametric model of errors}\label{sec:parametric_errors}
        SCIFYsim is primarily an end-to-end simulator designed to provide realistic end-product of the observation, but producing Monte-Carlo simulations for each of the relevant pointing directions and star magnitudes would require prohibitive amounts of processing power. So instead, we evaluate from a single Monte-Carlo time series, the different contributions to errors so they can be scaled independently.\par

        Let $F_{\mathrm{source}}$ be the flux total flux of a given source across the wavelength range of interest. We call $\mathbf{S}_{\mathrm{source}} = [S_{\mathrm{source, bin}}]$ the vector containing the spectrum of this source as collected on the spectral bins of the spectrograph. Leveraging linearity, to scale both signal and errors in this model with brightness of the source and integration time, we write:
        \begin{equation}
            \mathbf{S}_{\mathrm{source}} = F_{\mathrm{source}}  \int_{\mathrm{DIT}} \mathbf{r}(\mathbf{s}_{\mathrm{\mathrm{source}}}) \mathrm{d}t
        \end{equation}
        where $\mathbf{s}_{\mathrm{\mathrm{source}}}$ is the spectral distribution of the source and $\mathbf{r}$ is the instantaneous instrumental response function which acts independently on all spectral bins. It depends, as described earlier, on the source location of the field of view, the projection of the collecting array, the phase and amplitude input errors and the combiner transfer matrix. For spatially incoherent sources, like the thermal background, it is constant.\par
        
        Here we intend to account for the covariance in the errors. The covariance matrix of the error in a single DIT contains different contributions:
        \begin{equation}
            \boldsymbol{\Sigma}_{\mathrm{DIT}} = \boldsymbol{\Sigma}_{\mathrm{RON}} + \boldsymbol{\Sigma}_{\mathrm{photon}} + \boldsymbol{\Sigma}_{\mathrm{instrumental}},
        \end{equation}
        where $\boldsymbol{\Sigma}_{\mathrm{RON}}$, the covariance of the readout noise, and $\boldsymbol{\Sigma}_{\mathrm{photon}}$, the covariance of the photon noise, are diagonal matrices carrying the variance of the respective contributions, but the covariance of the instrumental noise $\boldsymbol{\Sigma}_{\mathrm{instrumental}}$ is not diagonal. While it normally contains contributions of all the sources in the field of view, it can be simplified, as the incoherent background sources have only a static contribution\footnote{The variability of the background light is currently neglected by SCIFYsim for simplicity} and the planet luminosity is orders of magnitudes less luminous than the star\footnote{Instrumental noise on an off-axis source would be difficult to account for: correlated, non-Gaussian and variable depending on the position in the map.}, leaving only the fluctuations of the light from the central star. The effect of both injection errors and OPD errors produce variations of the starlight on the interferometric output across the different different wavelength channels in a deterministic manner. The amplitude of this effect is directly proportional to the luminosity of the host star:
        \begin{equation}
            \boldsymbol{\Sigma}_{\mathrm{instrumental}} \approx F_{\mathrm{star}}\boldsymbol{ \Theta}_{\varphi} .
        \end{equation}
        
        Here, we evaluate this normalized covariance matrix $\boldsymbol{ \Theta}_{\varphi}$ empirically through MC simulations on a reference star of flux $F_{\mathrm{ref}}$ giving the recorded signal $\mathbf{S}_{\mathrm{ref}}$  :
        \begin{equation}
            \boldsymbol{ \Theta}_{\varphi} =  \frac{1}{F_{\mathrm{ref}}} \mathrm{Cov}(\mathbf{S}_{\mathrm{ref, MC}}).
        \end{equation}
        
        The readout noise term can be computed as the sum of the variances of the $n_{\mathrm{pix}}$ pixels contributing to the differential signal. This total number includes respectively the number used along the spectral dispersion $n_{\mathrm{bin}}$, the number of outputs contributing $n_{\mathrm{out}}$ (typically 2), and the number of contributing polarization channels $n_{\mathrm{pol}}$ :
        \begin{equation}
            n_{\mathrm{pix}} = n_{\mathrm{bin}} \times n_{\mathrm{out}} \times n_{\mathrm{pol}}.
        \end{equation}
        
        The photon noise term $\boldsymbol{\Sigma}_{photon} = \mathrm{diag}(\boldsymbol{\sigma}_{photon}^2)$ follows a Poisson distribution and can be approximated to a normal distributions in the case of large number of photoelectrons. For each bin, the variance writes:
        \begin{equation}
            \sigma_{\mathrm{photon, bin}}^2 \approx n_{\mathrm{out}} t_{\mathrm{DIT}} (S_{\mathrm{thermal, bin}} + \langle S_{\mathrm{leakage, bin}} \rangle) 
        \end{equation}
        as a function of the thermal background $\mathbf{S}_{\mathrm{thermal}}$ and the stellar leakage $\mathbf{S}_{\mathrm{leakage}}$.
        
        In this first approach, we assume that measurements made over each DIT are statistically independent. For convenience in the processing we decompose the observing sequence into $n_{\mathrm{chunk}}$ chunks (observing blocks) for which the signal from the $n_{\mathrm{DIT}}$ measurements is averaged, resulting in a covariance of:
        \begin{equation}
            \boldsymbol{\Sigma}_{\mathrm{chunk}} = \frac{1}{\sqrt{n_{\mathrm{DIT}}}} \boldsymbol{\Sigma}_{\mathrm{DIT}}.
        \end{equation}

\begin{figure}
            \centering
            \includegraphics[width=0.45\textwidth]{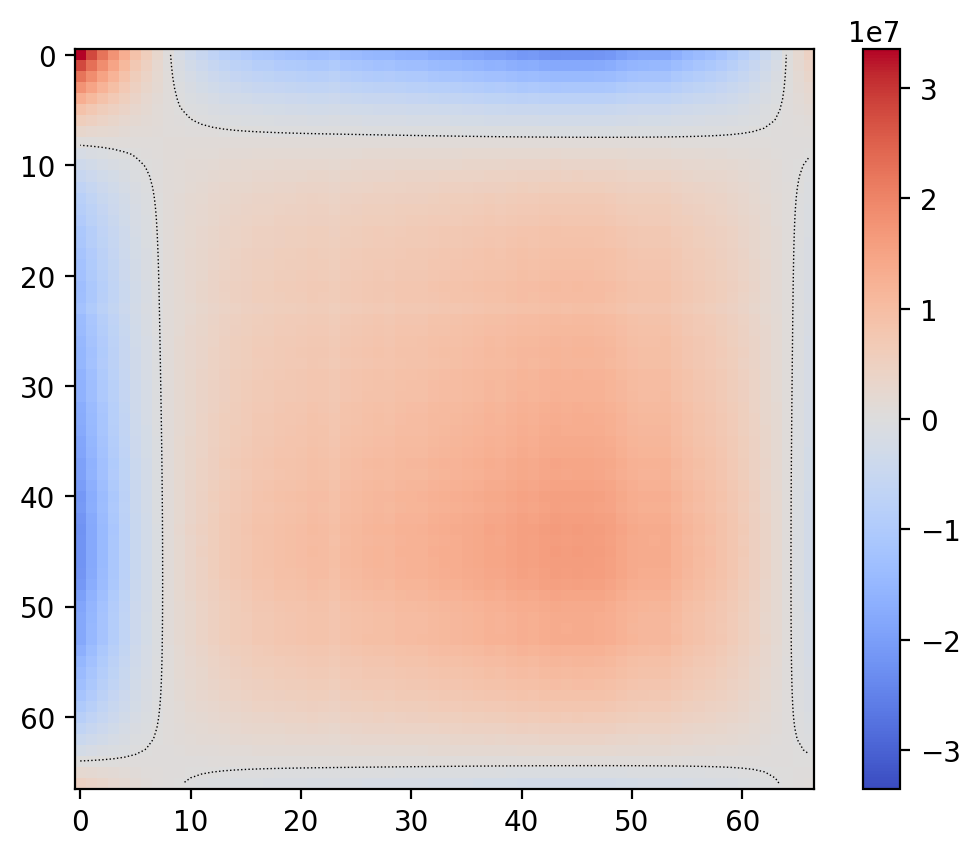}\par
            \includegraphics[width=0.45\textwidth]{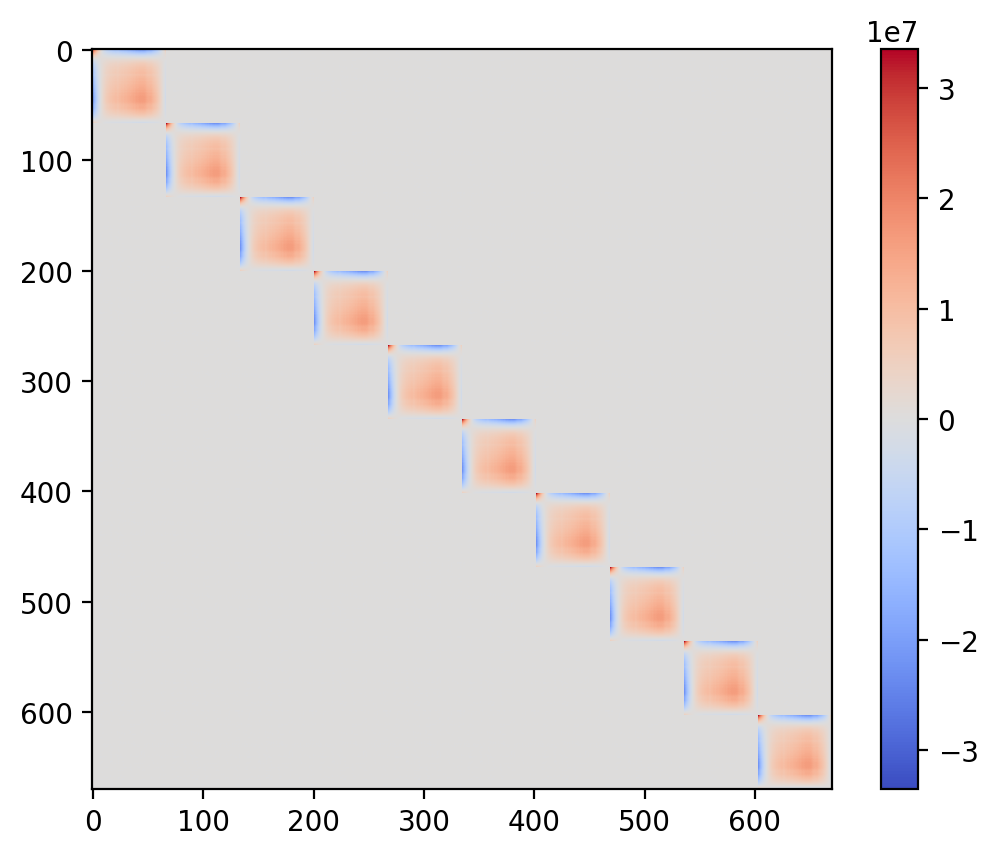}
            \caption{A color-mapped representation of the covariance matrix $\boldsymbol{\Sigma}_{\mathrm{chunk}}$ of the errors, for the case of a star of magnitude 4. The values of the diagonal, resulting from independent, pixel-level noise (photon noise and read-noise) are dwarfed by the correlated instrumental noise. The dotted line separates the positive and negative covariance. $\boldsymbol{\Sigma}$ is the block-diagonal matrix populated by the first one.}
            \label{fig:covariance}
        \end{figure}

        We combine the result over the full observing sequence into an single vector $\mathbf{z}$ by concatenation of the observable vectors of each of the observing blocks. Assuming the errors between the blocks are uncorrelated, the covariance matrix of $\boldsymbol{\Sigma}$ of $\mathbf{z}$ is a block-diagonal matrix containing the $\boldsymbol{\Sigma}_{\mathrm{chunk}}$ on the diagonal (see illustration in Fig. \ref{fig:covariance}).\par

\begin{figure}
            \centering
            \includegraphics[width=0.45\textwidth]{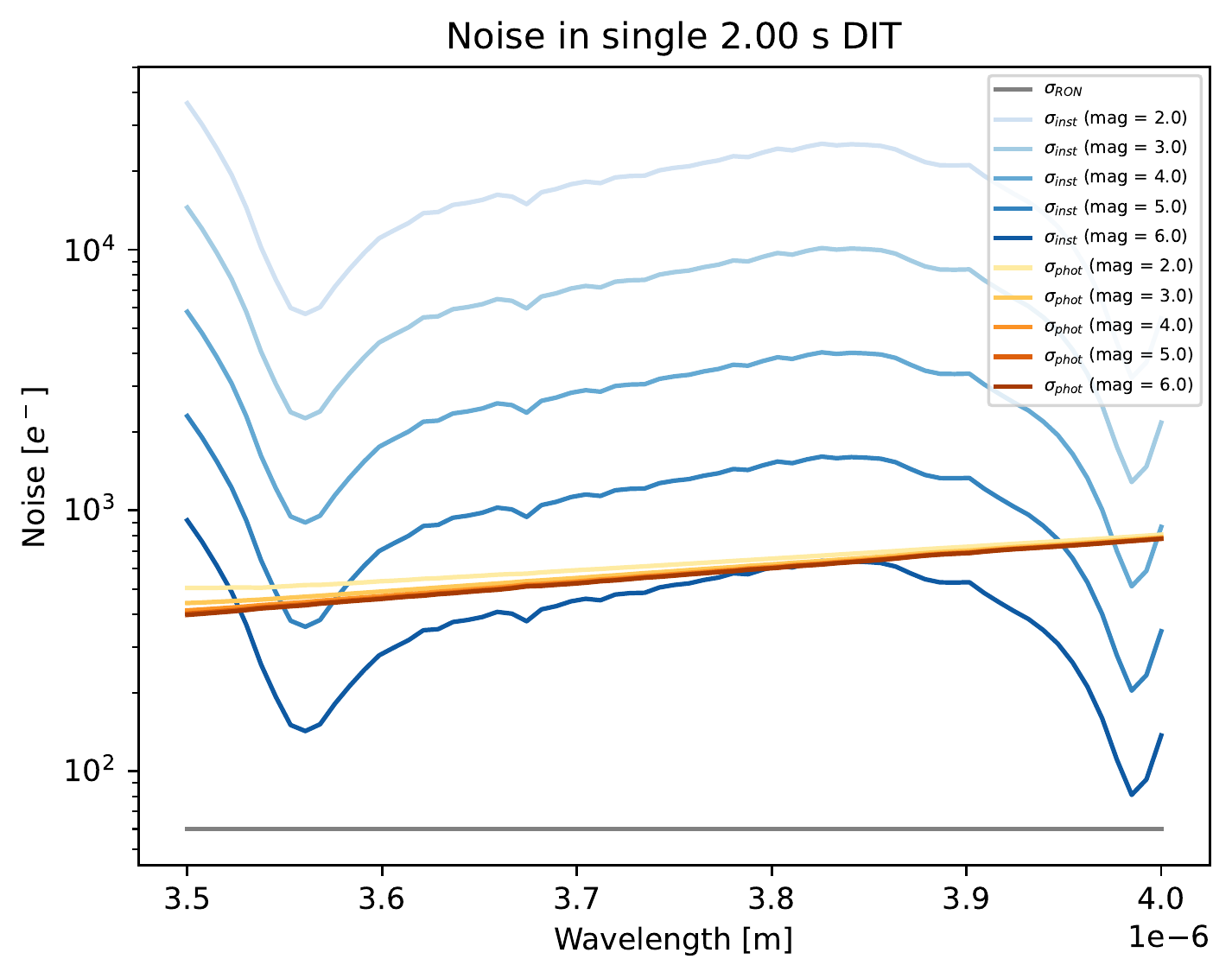}
            \caption{The standard deviation of the different error contributions for different star luminosity, as a function of wavelength. The read noise term is assumed constant. The photon noise term is dominated by the background at longer wavelength, but the effect of leakage light can be seen at the shorter wavelengths. The instrumental noise dominates for the brighter stars and follows the overall trend of Fig. \ref{fig:null_and_shape}, with small ripples caused by atmospheric absorption lines.}
            \label{fig:noise_model}
        \end{figure}

        Readout noise is added to each integration. To maximize the use of the dynamic range and minimize the added readout noise, $t_{\mathrm{DIT}}$ is chosen so that the flux on the dark outputs, here dominated by the static thermal flux, comes close to saturation:
        \begin{equation}
            \mathrm{Max}_{channels}(t_{\mathrm{DIT}}S_{\mathrm{thermal}}) \approx S_{\mathrm{full~well}}.
        \end{equation}
        The values used are detailed in Table \ref{tab:characteristics}. An example of this covariance for the case of a star at magnitude $L' = 4$ is displayed in Fig. \ref{fig:covariance}, and shows how the independent diagonal terms is swamped by the correlated instrumental noise. The relative scale of the different contributors to the errors are represented together in Fig. \ref{fig:noise_model}.\par
        
        Closer examination of the covariance reveals a property of the three spectral regions mentioned in Sect. \ref{sec:tuning}. The errors in the inner part of the spectrum ($3.57< \lambda<3.98~\mathrm{\mu m}$ are correlated together, as indicated with the red parts of the covariance matrix; but anti-correlated with the outer regions ($\lambda<3.57~\mathrm{\mu m}$ and $\lambda > 3.98~\mathrm{\mu m}$), as indicated by the blue parts of the matrix.

    \section{Statistical detection tests}\label{sec:tests_new}
        In order to evaluate the performance of the instrument for the detection of sources, we compute the detection limit of statistical tests. We use the tests described by \cite{Ceau2019b} for the detection in kernel-phases, but work here in a similar fashion.\par
        
        This is applied by considering the vector of observables that concatenates the different spectral channels of the instrument, and the different observations chunks in the session. The resulting vector has a length of $ n_{\mathrm{chunk}} \times n_{\mathrm{channels}} \times n_{\mathrm{outputs}} $. We neglect the temporal correlations between observation chunks. As a consequence, the resulting covariance matrix is block-diagonal, with square blocks of dimension $n_{\mathrm{channels}}$.\par
        
        Following the same principles, comparison between model and observation writes in the subspace of the differential null:
        \begin{equation}
            \mathbf{z} = \mathbf{z}^0 + \boldsymbol{\varepsilon}'
        \end{equation}
        where $\mathbf{z}^o$ is the theoretical signal produced by a model object. When the model matches reality, $\boldsymbol{\varepsilon}' \sim \mathcal{N}(0, \boldsymbol{\Sigma})$. In order to correctly express the likelihood of this with correlated errors, we use a change of variable through the same whitening transform used in \cite{Ceau2019b}.\par
        
        We produce a whitening matrix
        \begin{equation}
            \mathbf{W} = \boldsymbol{\Sigma}^{\frac{-1}{2}}
        \end{equation}
        to produce from the original vector of differential null $\mathbf{z}$ a new output vector $\mathbf{y} = \mathbf{W}.\mathbf{z}$ which is to be compared with a model signal $\mathbf{x} = \mathbf{W}.\mathbf{z}^o$. The new model-fitting equation writes
        \begin{equation}
            \mathbf{y} = \mathbf{x} + \boldsymbol{\varepsilon},
        \end{equation}
        with this time $\boldsymbol{\varepsilon}  \sim \mathcal{N}(0, \mathbf{I})$.
        
        The conclusions from that paper suggest that a more practical $T_B$ (generalized likelihood ratio) test would have an expected performance in lying in between those bounds. However, this case is not investigated here, as its sensitivity cannot be determined analytically. Instead, we examine the case of $T_E$, the energy detector, and $T_{NP}$ the Neyman-Pearson test. While $T_E$ offers a conservative estimate of the instrument performance, $T_{NP}$ provides a theoretical upper bound to detection performance.\par
        
        The amplitude of the signal of interest is computed so that it corresponds to the threshold $\xi$ of the test that satisfies both
        \begin{itemize}
            \item targeted $P_{FA}$, the false alarm probability, limiting the number of false detections,
            \item targeted $P_{Det}$, the detection probability, ensuring the sensitivity of the test.
        \end{itemize}
        This is solved in Sects. \ref{sec:TE} and \ref{sec:TNP}.
        
        \subsection{The energy detector test}\label{sec:TE}
            
            For the test $T_E$, we use a numerical method to solve the following system for $d = \mathbf{x}^T\mathbf{x}$ :
            \begin{equation}
                \begin{cases}
                        P_{FA}^{T_E} = 1 - \mathcal{F}_{\chi_p^2 (0)} (\xi) \\
                        P_{Det}^{T_E} = 1 - \mathcal{F}_{\chi_p^2 (b)} (\xi)
                \end{cases},
            \end{equation}
            where $\mathcal{F}_{\chi_p^2 (x)}$ is the cumulative distribution function (CDF) of the $\chi_2$ distribution with $p$ degrees of freedom in $x$.\par
            
            With the linearity of the system, for a given off-axis relative position, observing sources of flux $F$ and $F_0$ gives observable vectors of flux $\mathbf{x}$ and $\mathbf{x}_0$ respectively, so that:
            \begin{equation}
                \mathbf{x} = \frac{F}{F_0}\mathbf{x}_0 .
            \end{equation}
            
            As a consequence, the flux of a star to solve this system writes:
            \begin{equation}
                F = F_0 \frac{d}{\mathbf{x}_0^T\mathbf{x}_0},
            \end{equation}
            with $\mathbf{x}_0$ the value of the differential map, computed for a flux $F_0$.

        \subsection{The Neyman-Pearson test}\label{sec:TNP}
    
            For the Neyman-Pearson test, the law can be inverted analytically to obtain the sensitivity for a given false-alarm rate. This amounts to solving equation (17) in \cite{Ceau2019b} for $\mathbf{x}^T\mathbf{x}$. Starting from their equation (16), we write:
            \begin{align}
                \begin{cases}
                    \frac{\xi - \mathbf{x}^T\mathbf{x}}{\sqrt{\mathbf{x}^T\mathbf{x}}} = \mathcal{F}^{-1}_{\mathcal{N}(\mu=\mathbf{x}^T\mathbf{x},\, \sigma^2=\mathbf{x}^T\mathbf{x})}\Big( 1-P_{Det}  \Big) \\
                    \frac{\xi}{\sqrt{\mathbf{x}^T\mathbf{x}}} = \mathcal{F}^{-1}_{\mathcal{N}(\mu=0,\, \sigma^2=\mathbf{x}^T\mathbf{x})} \Big( 1-P_{FA} \Big) 
                \end{cases},
            \end{align}
            where $\mathcal{F}^{-1}_{\mathcal{N}(\mu=0,\, \sigma^2=\mathbf{x}^T\mathbf{x})}$ is the inverse of the cumulative distribution function of the normal distribution centered on zero of of variance $\mathbf{x}^T\mathbf{x}$.
            % Note in scipy (and numpy), scale is the standard deviation \sigma
            % Note also that F^-1 = CDF^-1 is called \mathrm{PPF} in scipy.stats
            From substitution, one can isolate the term representing the amplitude of the signal of interest:
            \begin{align}
                \sqrt{\mathbf{x}^T\mathbf{x}} = \mathcal{F}^{-1}_{\mathcal{N}(\mu=0,\, \sigma^2=\mathbf{x}^T\mathbf{x})} \Big( 1-P_{FA} \Big)\; - \nonumber\\
                    \mathcal{F}^{-1}_{\mathcal{N}(\mu=\mathbf{x}^T\mathbf{x},\, \sigma^2=\mathbf{x}^T\mathbf{x})}\Big( 1-P_{Det}  \Big)  .
            \end{align}
            This signal amplitude is proportional to the luminosity of the off-axis source to detect and depends on its position in the field of view and its spectrum. After computing the signal $\mathbf{x}_0$ obtained from a reference source of luminosity $F_0$ in each point of a map, one can rely on this linearity:
            \begin{equation}
                \frac{\sqrt{\mathbf{x}^T\mathbf{x}}}{F} = \frac{\sqrt{\mathbf{x}_0^T\mathbf{x}_0}}{F_0} 
            \end{equation}
            to compute the limit luminosity $F$ satisfying the criteria for the test:
            \begin{align}
                F = \frac{F_0}{\sqrt{\mathbf{x}_0^T\mathbf{x}_0}}  \Bigg(  \mathcal{F}^{-1}_{\mathcal{N}(\mu=0,\, \sigma^2=\mathbf{x}^T\mathbf{x})} \Big( 1-P_{FA} \Big)\; - \nonumber\\
                    \mathcal{F}^{-1}_{\mathcal{N}(\mu=\mathbf{x}^T\mathbf{x},\, \sigma^2=\mathbf{x}^T\mathbf{x})}\Big( 1-P_{Det}  \Big)   \Bigg) .
            \end{align}
            
            This sensitivity value will be offset by the spectral energy distribution of both the planet or the star and will be different from the model. For simplicity, we use here a flat spectrum for the planet, which is a good approximation in the L' band for young giant planets at temperatures around $1000~K$. Note also that deviations from the Gaussian hypothesis mentioned in Sect. \ref{sec:errors} will lead to biases in the actual false alarm rate. A large kurtosis in particular will lead to under-estimating $P_{FA}$. More thorough Monte-Carlo simulations may later be used to refine these thresholds by measuring empirically the distribution of the final test statistic.
            
\section{Application and results}\label{sec:application}
    \subsection{Example application case}\label{sec:case}
        NOTT is designed to operate as part of the Asgard suite with HEIMDALLR as high precision fringe tracker and the Baldr as wavefront sensor (based on Zernike wavefront sensor) which are currently under development. Here, we adjust the performance of wavefront control to match existing systems available at the Paranal observatory. We use the measured performance of the GRAVITY fringe tracker, and adjust the performance of adaptive optics to match the tip-tilt performance of NAOMI for the ATs with $25~\mathrm{mas}$ RMS \citep{Woillez2019} and SPHERE for the future GPAO system with $3~\mathrm{mas}$ RMS (Le Bouquin, private communication). The spatial sampling of the pupil is also matched with four and fourty control elements across.\par
        
\begin{figure}
            \centering
            \includegraphics[width=0.45\textwidth]{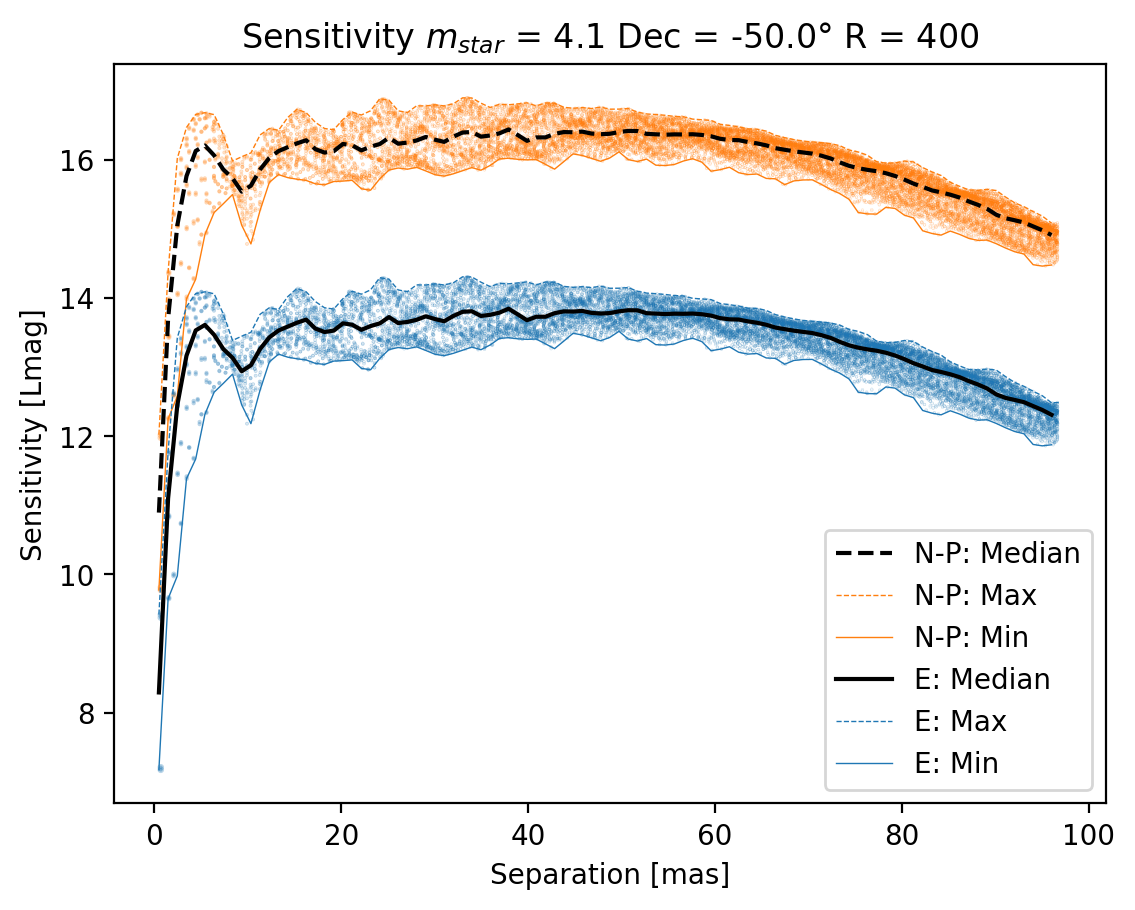}
            \caption{Radial scatter plot representation of each pixel of the sensitivity map. The map shows in blue the test $T_E$ and in orange the Neyman-Pearson test $T_{N-P}$. Bounds of min, max and median are also plotted. The two test show very similar behavior except for their different sensitivity.}
            \label{fig:sensitivity_cloud}
        \end{figure}

\begin{figure}
            \centering
            \includegraphics[width=0.45\textwidth]{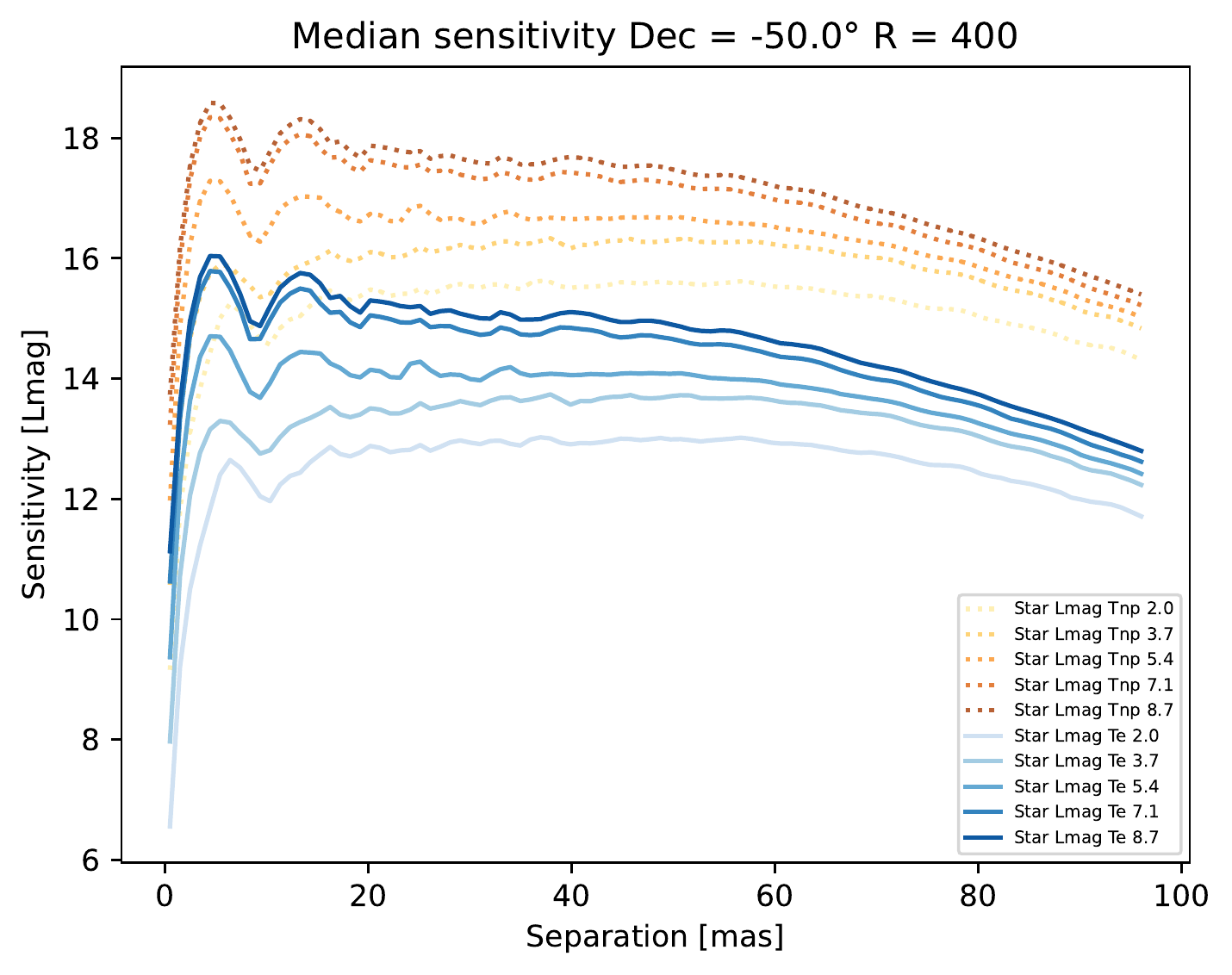}
            \caption{Aggregate of the median plots from Fig. \ref{fig:sensitivity_cloud} for various stellar magnitude and for the two tests. The dashed vertical line represents a separation of $2\lambda/D$ for the ELT.}
            \label{fig:sensitivity_curves}
        \end{figure}

\begin{figure}
            \centering
            \includegraphics[width=0.45\textwidth]{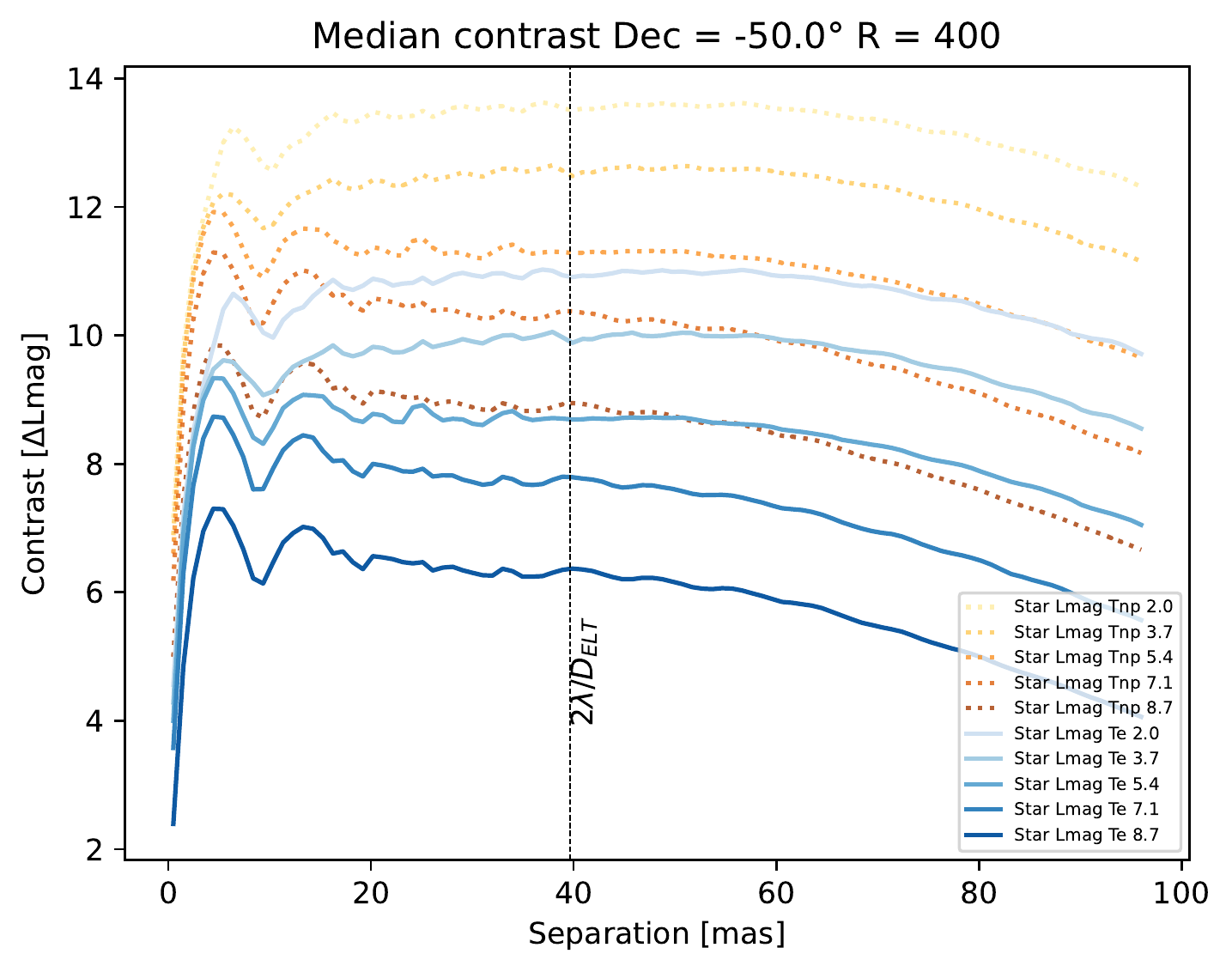}
            \caption{Aggregate of the median plots from Fig. \ref{fig:sensitivity_cloud} shown in contrasts for the two tests.}
            \label{fig:contrast_curves}
        \end{figure}

        The observed star is modeled after Gl 86 A, a nearby K dwarf of $L' \approx 4.1$ located at a declination of around $-50$ deg. Across the different curves in Figs. \ref{fig:sensitivity_curves} and \ref{fig:contrast_curves}, the angular diameter and temperature of the star are kept constant, and only magnitude is adjusted. We assume a sequence of 20 observation blocks equally distributed over six hours around its passage of the meridian. Total integration time on target is 3h (overheads are 50\%), yielding 180 DITs per block. Sensitivity is given for the detection of a planet with a flat spectrum for $P_{FA} = 0.01$ and $P_{Det} = 0.9$. The relevant parameters for Figs. \ref{fig:sensitivity_map} and \ref{fig:sensitivity_cloud} are given in Table \ref{tab:characteristics}.
        
        \begin{table}[]
            \centering
            \begin{tabular}{c c}
                \hline \hline
                \multicolumn{2}{c}{Astrophysical parameters}\\
                \hline
                $T_{\mathrm{star}}$ & $5263 ~\mathrm{K}$ \\
                $R_{\mathrm{star}}$ & $0.77 ~\mathrm{R_{\astrosun}}$\\
                $d$                 & $10.8 ~\mathrm{pc}$\\
                Magnitude L'        & $ 4.1$ \\
                Star spectrum model & Black body \\
                Planet spectrum     & Flat \\
                \hline
                \multicolumn{2}{c}{Instrumental parameters}\\
                \hline
                Collectors & UTs \\
                Telescope input order & 0, 2, 1, 3\\
                Spectral resolution R & $\approx 400$\\
                DIT & 3 s \\
                Combiner and spectro. temperature & $100 ~\mathrm{K}$ \\
                Detector temperature & $40 ~\mathrm{K}$\\
                Readout noise & $15 ~e^-$\\
                Dark current & $0.02 ~e^-s^{-1}\mathrm{pix}^{-1}$ \\
                \hline
            \end{tabular}
            \caption{Observation parameters for the example sensitivity map}
            \label{tab:characteristics}
        \end{table}

%    star_radius = 0.81                     # R_sun    Radius of the star
%star_distance = 3.6                    # pc       Distance to the star
%star_temperature = 5375.   
        
    \subsection{Results}\label{sec:results}
        The simulation of several hundred frames provides the covariance of the observables as a function of star magnitude. Using the transmission map of the instrument and the equations of section 4.3, this covariance is used to produce the sensitivity map for each of the detection tests for any star magnitude. Both look alike with a different scale, and the map for $T_E$ is shown in Fig. \ref{fig:sensitivity_map}. The way this map is affected by the position on sky is discussed in appendix \ref{app:variation}.
        
        \begin{figure}
            \centering
            \includegraphics[width=0.45\textwidth]{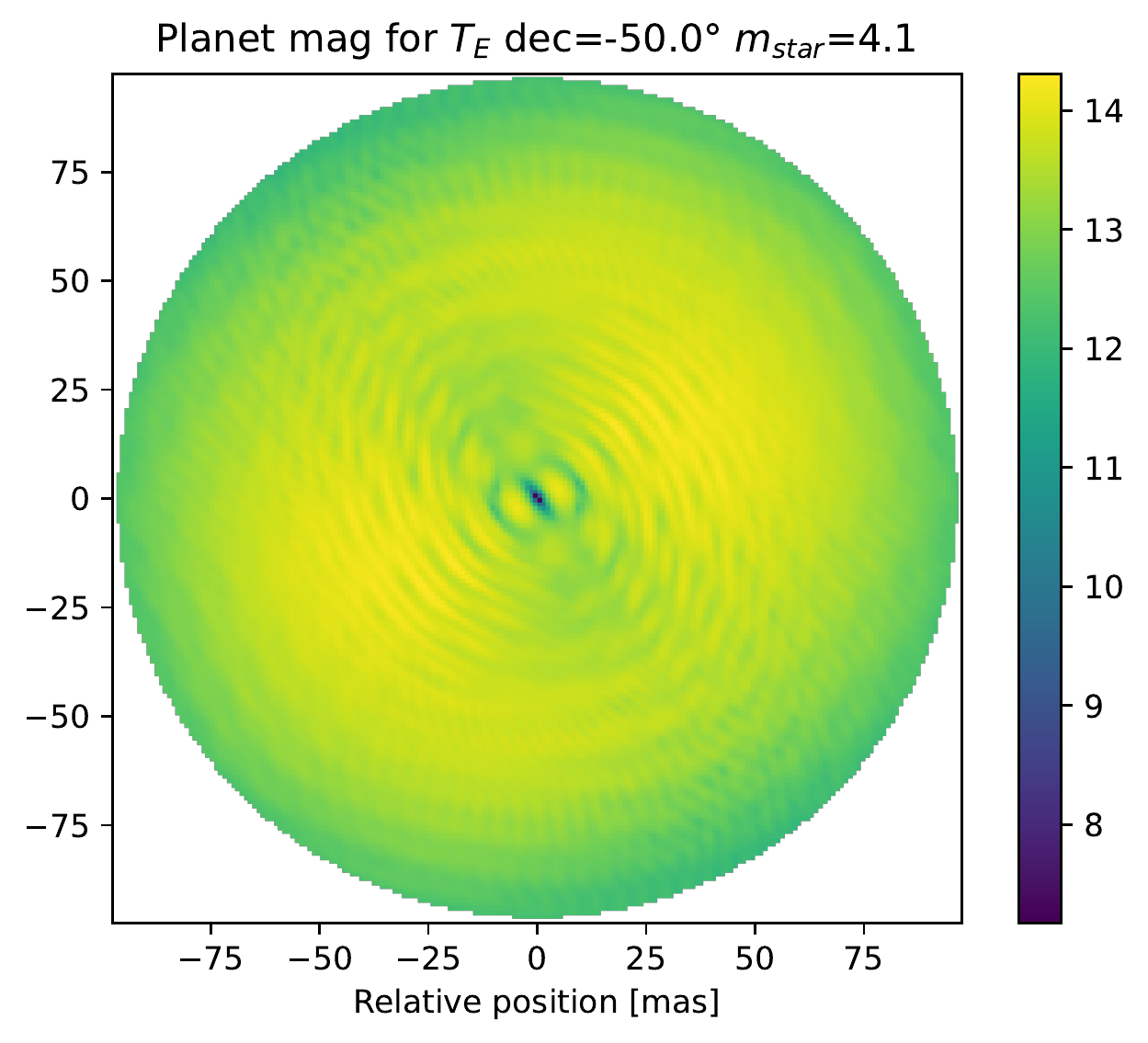}
            \caption{Sensitivity map showing the maximum magnitude of a detectable companion with the energy detector test with $P_{FA} = 0.01$ and $P_{Det} = 0.9$, at a declination of -50.0 degrees and around a star of $L = 4.1$. The map evolves in shape for different declinations, but not significantly in overall sensitivity.}
            \label{fig:sensitivity_map}
        \end{figure}
        
        At the smallest separations, the lobes of the transmission map at different wavelengths overlap, creating a first broadband lobe. This lobe offers the most transmission and therefore gives the most sensitivity and performance, up to $\approx 1$ magnitude above the field. However, on the brighter stars, when the instrumental noise dominates, performance decreases in the first lobes compared to the outer regions. In the first peak, a flat-spectrum planet would produce broad spectrum signal in the data, consistent with highly correlated instrumental noise. However at larger separations, the signal of a flat-spectrum planet would appear as multiple positive and negative bumps in the differential null, distinct from the correlated instrumental noise.\par
        
        This shows the importance of accounting for the correlated instrumental noise which allows us to still salvage the signal that does not correlate in the same way as the data.

\section{Discussion}\label{sec:discussion}
    
    The dependency we have highlighted between the high contrast sensitivity and the stellar luminosity was examined assuming the performance of wavefront control remains constant. In practice it will also vary with the luminosity of the star, its spectral type, the atmospheric conditions and the airmass, but a finer study of this performance is beyond the scope of this work, and will concern the wavefront control provided by the HEIMDALLR \citep{Ireland2018} for fringe tracking and Baldr for refined adaptive optics which should yield improved performance.\par
    
    The atmospheric dispersion is not taken into account in this first round of simulations. The static part of its effect will be compensated using a combination of air and ZnSe additional optical path via forward modeling. The dynamical part of this dispersive effect, often referred to as water vapor seeing, is not taken into account here, and its impact on the performance will be the topic of future work.\par
    
    The effect of transverse atmospheric dispersion is not expected to be a large contributor of errors, with an effect of the order of 3~mas in pointing. Nevertheless, this will be implemented in the simulator in the near future, so as to correctly model its effect on the correlations of the nulled outputs.\par

    The effect of polarization is not taken into account in this work. Characterization of the polarization in VLTI is currently incomplete \citep{LeBouquin2008, Lazareff2014a}, but most of the problems will be handled in the instrument with a Wollaston prism in the spectrograph and birefringence compensation plates in each beam \citep{Lazareff2012, LeBouquin2011}.\par
    
    The confidence estimates offered by the detection test suggested in Sect. \ref{sec:tests_new} are contingent upon the Gaussian distribution of the differential nulled output. In practice, with the level of wavefront correction used as a baseline here, a deviation from this hypothesis is measurable with a Shapiro-Wilk test on a few hundred Monte-Carlo samples. The temporal correlations in the measurement affect both our estimation of the amplitude of errors on the mean, and its distribution. This does not invalidate the usefulness of the likelihood ratio framework as a whole. Accounting for these effects is beyond the scope of this work. Future efforts will include frame selection relying on metrology and the measurements of the photometric outputs to better constrain the distribution by removing some outliers. Adapting the tests to use a distribution that better model the actual distribution of the differential output is also possible.\par
    
    Here we have used a fixed stellar spectrum and a fixed flat planet spectrum in order to produce results as a function of simple luminosity. Deviations of the planet spectrum impacts the signal level $\sqrt{\mathbf{x}^T\mathbf{x}}$. Deviation on the stellar spectrum impacts the covariance matrix of errors, which therefore impacts the sensitivity result through the whitening matrix.
    
\section{Conclusion}\label{sec:conclusion}
    SCIFYsim offers a new framework to model integrated optics beam combiner for interferometry, both in aperture-masking, and long-baseline configurations. It models the complex effects of fringe-tracking residuals injection into the waveguides, as well as the chromatic amplitude and phase aberrations intrinsic to the integrated optics component. This emphasis on instrumental errors is suitable for the demanding case of nulling interferometry, and allows for the evaluation of the covariance of the errors on the outputs. Its flexibility makes it ready to work with new sophisticated combiner architectures such as kernel-nulling.\par
    
    We show how the knowledge of the covariance of errors can be used to determine the performance of the instrument in a realistic scenario. The approach takes into account the rotation of the Earth, the combination of all spectral channels, and the covariance of the instrumental errors. To our knowledge, this is a first for nulling interferometry observations. The approach provides upper and lower bounds for the sensitivity of matched subspace detectors to single companions around the central object for optimal detection tests depending on the initial hypotheses on spectrum and position.\par

    Intitial results show performance in contrast detection comparable to ground-based coronagraphy and to GRAVITY in off-axis mode, down to a few $\lambda / D $ of the single dish telescopes. Our simulations show that nulling interferometry can attain this performance down to an angular separation of $\lambda /B$ for long-baseline interferometers, i.e. beyond the reach of extremely large telescopes. Future work will incorporate atmospheric dispersion and its compensation into the simulator, and elaborate a target selection strategy to maximize the scientific yield.
    %Initial results show performance compatible to the detection of young giant planets with contrast of more than $2.10^4$ in L' with dispersion $R \approx 400$ around bright stars. Such performance is comparable to what has been achieved with ground-based coronagraphy and with GRAVITY in off-axis mode, down to a a few $\lambda / D $ of the single dish telescopes. Our simulations show that nulling interferometry can attain this performance down to $\lambda /B$ of long-baseline interferometers, reaching angular separations beyond the reach of extremely large telescopes. It also demonstrates a strategy to handle the spectral correlations in the nulling data, and the rotation of the earth in a common framework.\par
    
    The tools were designed to be flexible and could be used to investigate the on-sky performance of other single-mode beam combiners, such as GRAVITY, GLINT, or future missions like LIFE.

\begin{acknowledgements}
    SCIFY has received funding from the European Research Council (ERC) under the European Union's Horizon 2020 research and innovation program (grant agreement CoG - 866070). This project has received funding from the European Union’s Horizon 2020 research and innovation programme under grant agreement No 101004719. The authors wish to thank Stephane Lagarde, Fatmé Allouche, and Philippe Berio for their helpful suggestions regarding the design of MATISSE.
\end{acknowledgements}

\bibliographystyle{aa}
\bibliography{scifysim}

\begin{appendix}
    \section{Variation of the sensitivity maps with configuration and sky position}\label{app:variation}
        The sensitivity map is affected by multiple parameters of the pupil. The first is the position of the collecting telescopes, through the length and orientation of the baselines it offers. Here we will use the position of the four UTs as an example.\par
        
        The second is the diameter of individual apertures. Through the spatial filtering of the single-mode waveguide, they dictate the field of view and its progressive falloff at larger separations.\par
        
\begin{figure}
            \centering
            \includegraphics[width=0.45\textwidth]{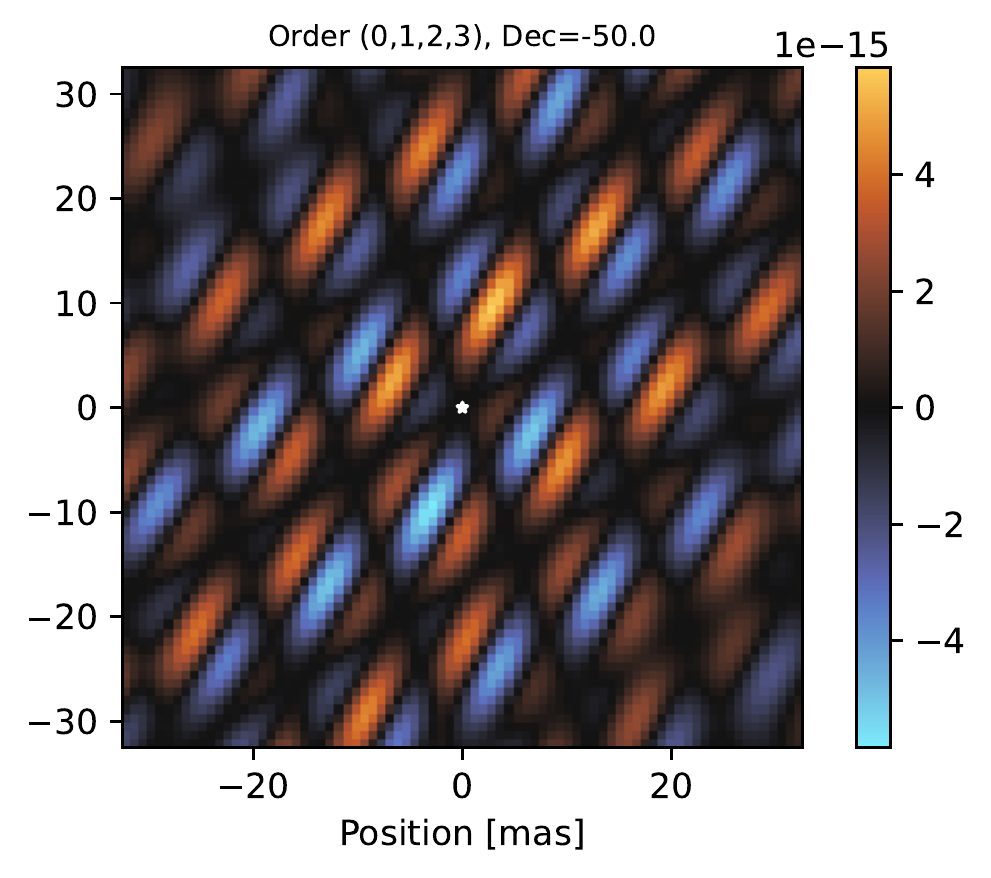}\par
            \includegraphics[width=0.45\textwidth]{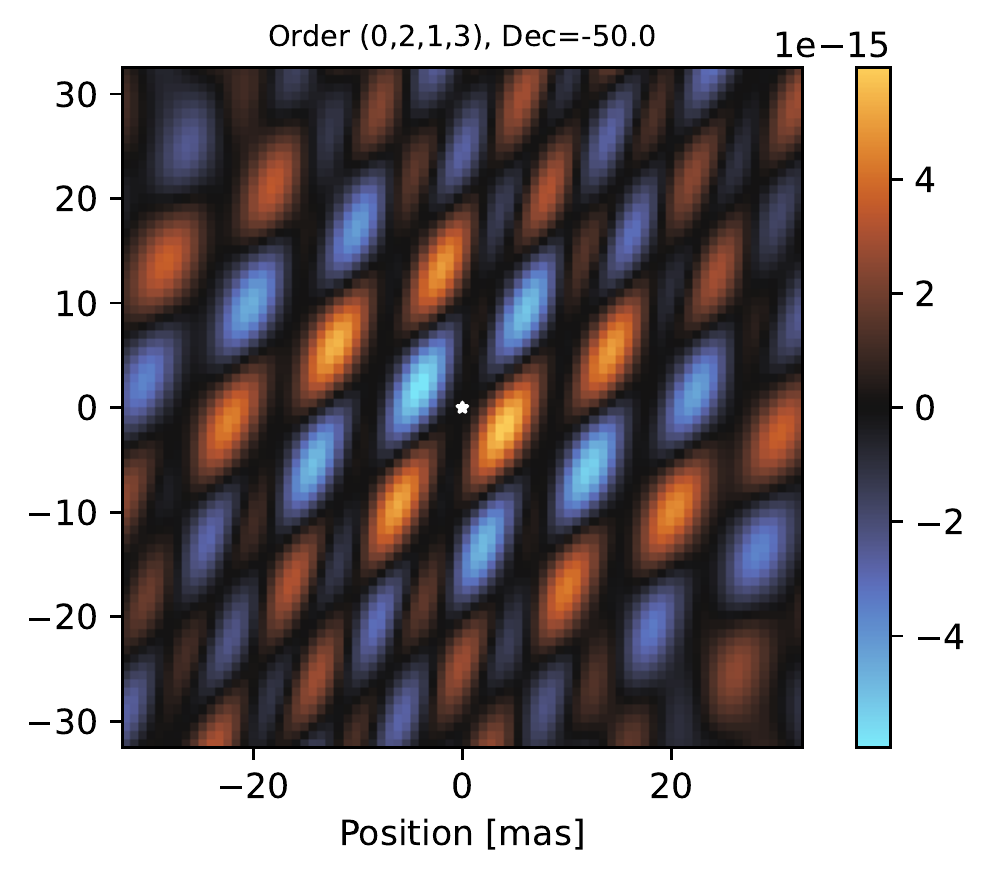}\par
            \includegraphics[width=0.45\textwidth]{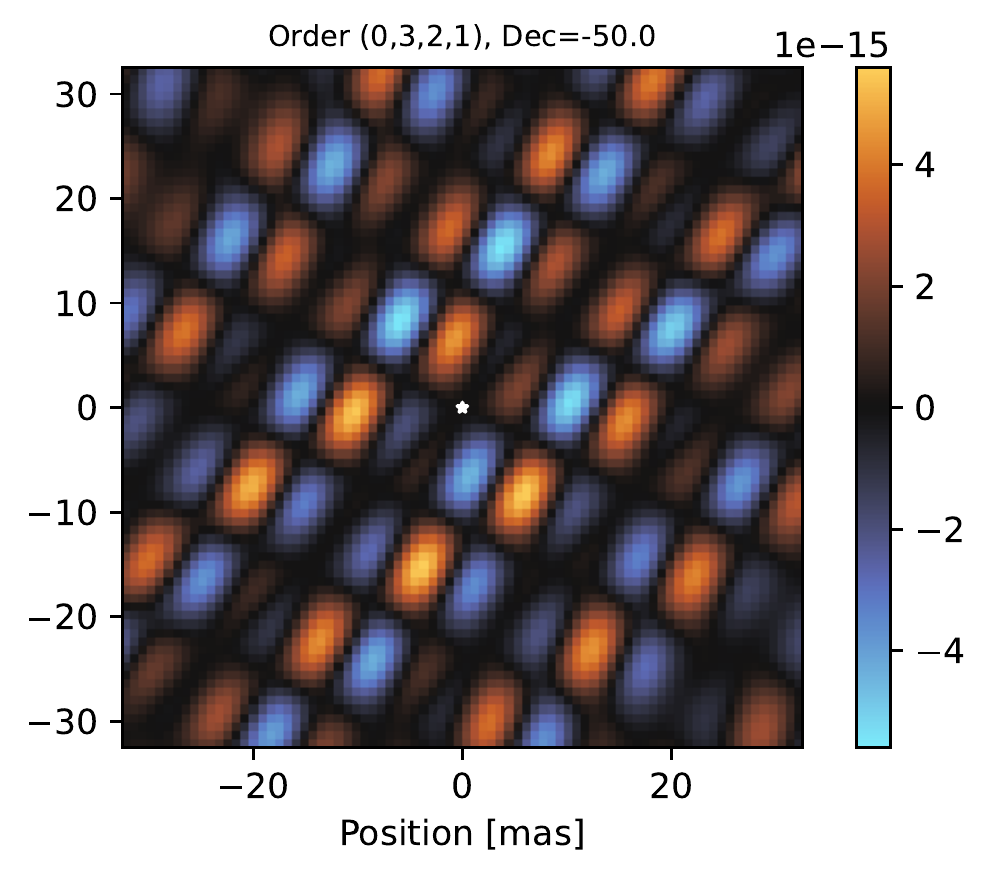}\par
            \caption{The raw transmission maps of the differential null for three different input orders and the same sky pointing, on the meridian for a target at Dec $-50.0$. The signal is summed over all spectral channels. Unlike the four-telescope kernel-nuller, but NOTT can only obtain one at a time. The values take into account the transmission of the whole instrument, (including quantum efficiency etc.) boiling down to units of étendue $[\mathrm{sr}.\mathrm{m}^2]$, interpreted as the equivalent collecting capacity with regard to each elementary patch of sky (pixel).}
            \label{fig:raw_maps}
        \end{figure}

        The third is the permutation of the collecting telescopes with respect to the inputs of the instrument. The double Bracewell configuration of NOTT offers only one pair of dark outputs, which corresponds to one of the kernel pairs of the VIKiNG solution proposed by \cite{Martinache2018}. However one can produce the three different configurations of the VIKiNG nuller by observing with permutations of the input telescopes (Fig. \ref{fig:raw_maps} for one given sky pointing).\par
        
        The fourth is the pointing in the sky which changes the projected geometry of the array. Throughout an observing session, the rotation of the earth between the different chunks of the sequence changes this geometry by a projection into the plane orthogonal to the line of sight, effectively stretching and rotating the maps of Fig. \ref{fig:raw_maps}. Within the parameters used here, the effect of airmass on throughput and added photon noise are negligible, as this case is not sensitivity limited. The effect of airmass on the performance of adaptive optics, fiber coupling and fringe tracking is not taken into account here.
        
\begin{figure}
            \centering
            \includegraphics[width=0.45\textwidth]{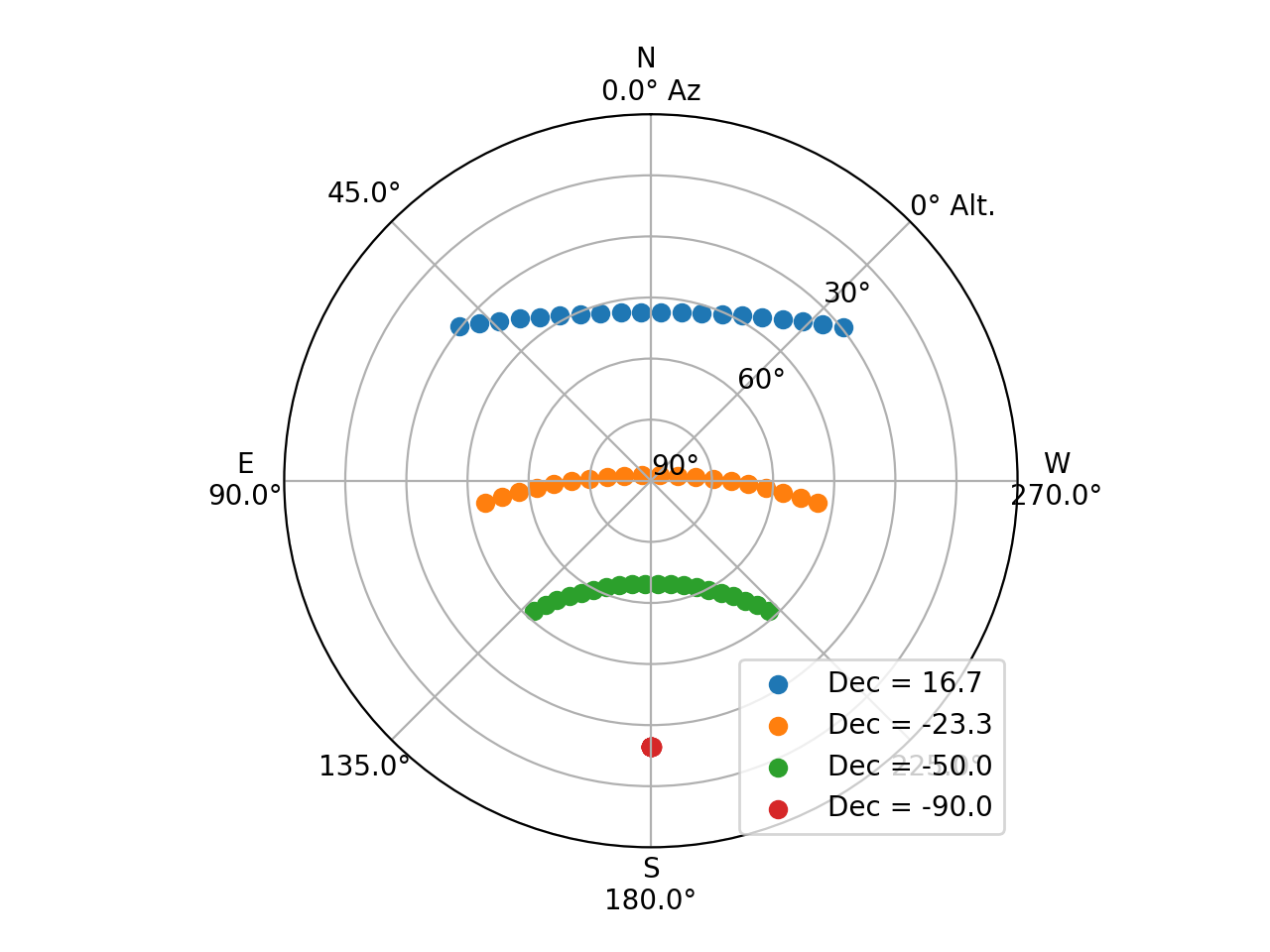}
            \caption{A map showing altitude and azimuth of the series of pointings combined to establish the maps in Fig. \ref{fig:various}.}
            \label{fig:trans_maps}
        \end{figure}

\begin{figure*}
            \centering
            \includegraphics[width=0.94\textwidth]{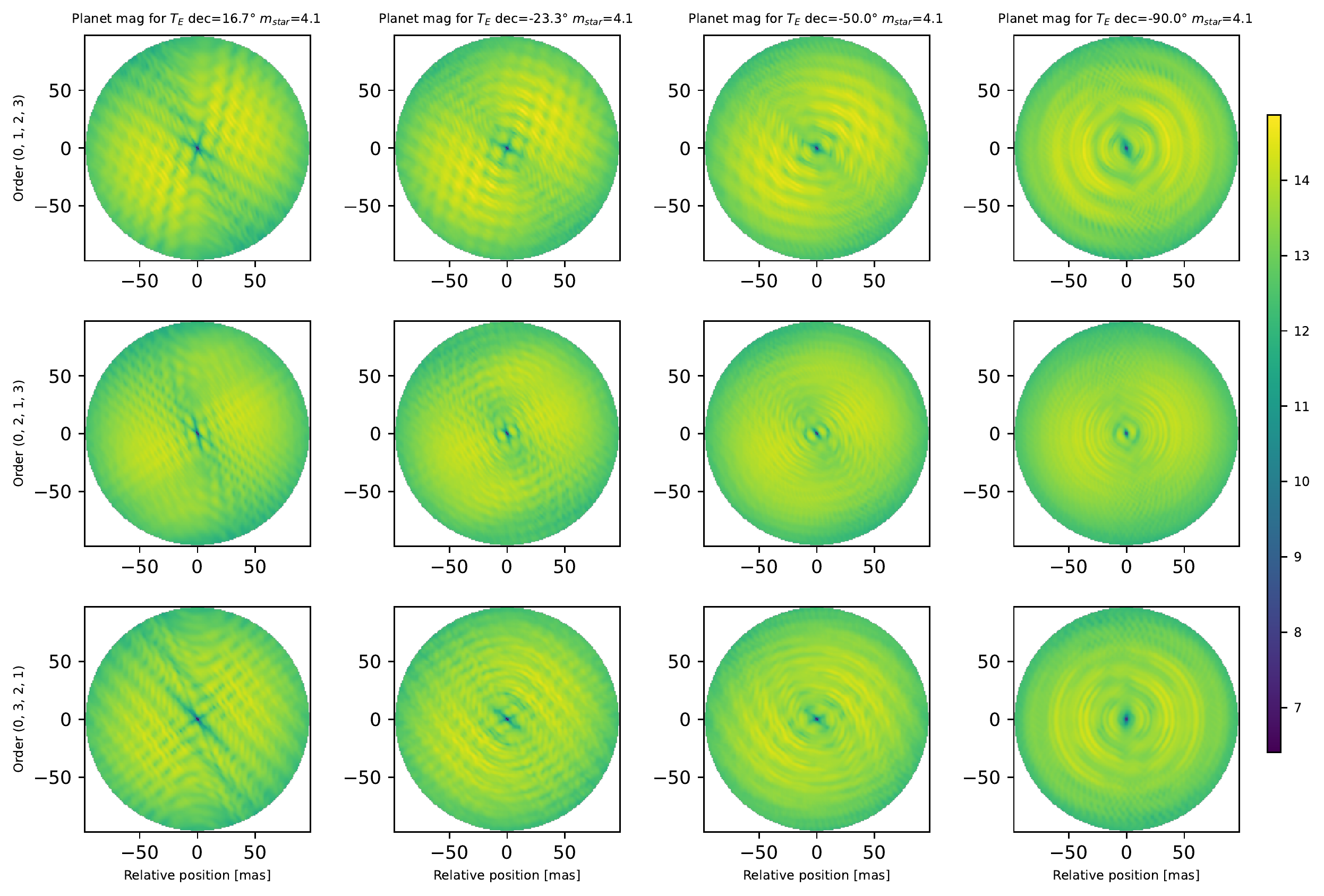}
            \caption{Sensitivity around a star of fixed magnitude, for different declinations and input order. The distribution of the sensitivity changes significantly, but the overall performance is preserved. The order (0, 2, 1, 3) seems to offer better sensitivity at the smaller separations, owing to the slightly tighter position of the first inner lobes of the transmission map if Fig. \ref{fig:trans_maps}.}
            \label{fig:various}
        \end{figure*}

        The overall sensitivity compounded throughout the session is affected. For comparison, we plot side-by-side the sensitivity map for targets of different declinations observed in the same six-hours series of 20 chunks spread throughout the passage of the meridian (see Fig. \ref{fig:trans_maps}). These maps are shown in Fig. \ref{fig:various} and are all distinct in the location of higher and lower sensitivity. The various sensitivity lobes are spread-out in a bow-like patterns, while the contribution of the different wavelenghts contribute to a radial blurring at some of the larger separations.
        
        For a blind search, one would argue that the overall sensitivity across the field of view appears equivalent, although some retain some better sensitivity around the first few mas around the host star, as has been shown in the first yield estimations by \cite{Dandumont2022a}. However for a targeted characterization observation with prior knowledge of the the position of the planet, this would help to pick a configuration maximizing the S/N. Even astrometry information, such as gathered by GAIA \citep{kervella2019, Kervella2022} can offer hints on the position angle of a suspected planet by the accelerations excess measured, and could prompt the use of a particular beam permutation.\par

\end{appendix}

\end{document}